\documentclass[11pt]{amsart}

\usepackage{graphicx,amsmath,amssymb,amsthm,bbm,url}
\usepackage{algorithm}
\usepackage{algorithmic}
\usepackage{fullpage}
\usepackage{color}
\usepackage{bm}
\usepackage{hyperref}
\usepackage{subcaption}
\usepackage{comment}
\usepackage{multirow}
\usepackage{wrapfig}
\usepackage{picins}

\newtheorem{lem}{Lemma}

\newtheorem{examp}{Example}

\begin{document}
\pagestyle{plain}

\title{Extension of PCA to higher order data structures:  An Introduction to Tensors, Tensor Decompositions, and Tensor PCA}\thanks{\noindent  A. Zare (zareali@msu.edu) is with the Department of Computational Mathematics, Science and Engineering (CMSE), Michigan State University, East Lansing, MI, 48824, USA.\\
\indent A. Ozdemir (ozdemira@egr.msu.edu), and S. Aviyente (aviyente@egr.msu.edu) are with the Electrical and Computer Engineering, Michigan State University, East Lansing, MI, 48824, USA.\\
\indent Mark A. Iwen (markiwen@math.msu.edu) is with the Department of Mathematics, and the Department of Computational Mathematics, Science and Engineering (CMSE), Michigan State University, East Lansing, MI, 48824, USA.\\
\indent This work was supported in part by NSF CCF-1615489.}
\author{Ali Zare, Alp Ozdemir, Mark A. Iwen, and Selin Aviyente}

\begin{abstract}
The widespread use of multisensor technology and the emergence of big data sets have brought the necessity to develop more versatile tools to represent higher-order data with multiple aspects and high dimensionality. Data in the form of multidimensional arrays, also referred to as tensors, arises in a variety of applications including chemometrics, hyperspectral imaging, high resolution videos, neuroimaging, biometrics, and social network analysis.
Early multiway data analysis approaches reformatted such tensor data as large vectors or matrices and then resorted to dimensionality reduction methods developed for classical two-way analysis such as PCA.  However, one cannot discover hidden components within multiway data using conventional PCA. To this end, tensor decomposition methods which are flexible in the choice of the constraints and that extract more general latent components have been proposed. In this paper, we review the major tensor decomposition methods with a focus on problems targeted by classical PCA. In particular, we present tensor methods that aim to solve three important challenges typically addressed by PCA: dimensionality reduction, i.e. low-rank tensor approximation, supervised learning, i.e. learning linear subspaces for feature extraction, and robust low-rank tensor recovery. We also provide experimental results to compare different tensor models for both dimensionality reduction and supervised learning applications.
\end{abstract}

\maketitle
\thispagestyle{empty}


\section{Introduction}
Principal Component Analysis (PCA) is one of the oldest and widely used methods for dimensionality reduction in data science. The goal of PCA is to reduce the dimensionality of a data set, i.e.  extract low-dimensional subspaces from high dimensional data, while preserving as much variability as possible \cite{shlens2014tutorial}.  Over the past decades, thanks to its simple, non-parametric nature, PCA has been used as a descriptive and adaptive exploratory method on numerical data of various
types. Currently, PCA is commonly used to address three major problems in data science: 1) Dimensionality reduction for large and high dimensional data sets, i.e. low-rank subspace approximation \cite{cunningham2015linear}; 2) Subspace learning for machine learning applications  \cite{jiang2011linear}; and 3) Robust low-rank matrix recovery from missing samples or grossly corrupted data \cite{candes2011robust,wright2009robust}.  However, these advances have been mostly limited to vector or matrix type data despite the fact that continued advances in information and sensing technology have been making large-scale, multi-modal, and multi-relational datasets evermore commonplace.  Indeed, such multimodal data sets are now commonly encountered in a huge variety of applications including chemometrics \cite{wu2009multi}, hyperspectral imaging \cite{letexier2008nonorthogonal}, high resolution videos \cite{kim2009canonical}, neuroimaging (EEG, fMRI) \cite{miwakeichi2004decomposing}, biometrics \cite{tao2007general,fronthaler2008fingerprint} and social network analysis \cite{zhang2014cap,maruhashi2011multiaspectforensics}. When applied to these higher order data sets, standard vector and matrix models such as PCA have been shown to be inadequate at capturing the cross-couplings across the different modes and burdened by the increasing storage and computational costs \cite{cichocki2014era,cichocki2015tensor,sidiropoulos2017tensor}. Therefore, there is a growing need for PCA type methods that can learn from tensor data while respecting its inherent multi-modal structure for multilinear dimensionality reduction and subspace estimation.

The purpose of this survey article is to introduce those who are well familiar with PCA methods for vector type data to tensors with an eye toward discussing extensions of PCA and its variants for tensor type data. Although there are many excellent review articles, tutorials and book chapters written on tensor decomposition (e.g. \cite{cichocki2015tensor,sidiropoulos2017tensor,kolda2009tensor,acar2009unsupervised,cichocki2016tensor}), the focus of this review article is on extensions of PCA-based methods developed to address the current challenges listed above to tensors. For example, \cite{kolda2009tensor} provides the fundamental theory and methods for tensor decomposition/compression focusing on two particular models, PARAFAC and Tucker models, without much emphasis on tensor network topologies, supervised learning and numerical examples that contrast the different models. Similarly, \cite{sidiropoulos2017tensor} focuses on PARAFAC and Tucker models with a special emphasis on the uniqueness of the representations and computational algorithms to learn these models from real data. Cichocki et al. \cite{cichocki2016tensor}, on the other hand, mostly focus on tensor decomposition methods for big data applications, providing an in depth review of tensor networks and different network topologies. The current survey differs from these in two key ways. First, the focus of the current survey is to introduce methods that can address the three main challenges or application areas that are currently being targeted by PCA for tensor type data. Therefore, the current survey reviews dimensionality reduction and linear subspace learning methods for tensor type data as well as extensions of robust PCA to tensor type data. Second, while the current survey attempts to give a comprehensive and concise theoretical overview of different tensor structures and representations, it also provides experimental comparisons between different tensor structures for both dimensionality reduction and supervised learning applications.

In order to accomplish these goals, we review three main lines of research in tensor decompositions herein.  First, we present methods for tensor decomposition aimed at low-dimensional/low-rank approximation of higher order tensors.  Early multiway data analysis relied on reshaping tensor data as a matrix and resorted to classical matrix factorization methods. However, the matricization of tensor data cannot always capture the interactions and couplings across the different modes.  For this reason, extensions of two-way matrix analysis techniques such as PCA, SVD and non-negative matrix factorization were developed in order to better address the issue of dimensionality reduction in tensors.  After reviewing basic tensor algebra in Section~\ref{sec:Setup}, we then discuss these extensions in Section~\ref{sec:TensorDecomps}.  In particular, we review the major tensor representation models including the CANonical DECOMPosition (CANDECOMP), also known as PARAllel FACtor (PARAFAC) model, the Tucker, or multilinear singular value, decomposition \cite{Lathauwer_Multilinear_2000} and tensor networks, including the Tensor-Train, Hierarchical Tucker and other major topologies. These methods are discussed with respect to their dimensionality reduction capabilities, uniqueness and storage requirements. Section~\ref{sec:TensorDecomps} concludes with an empirical comparison of several tensor decomposition methods' ability to compress several example datasets. Second, in Section~\ref{sec:TensorPCA}, we summarize extensions of PCA and linear subspace learning methods in the context of tensors. These include Multilinear Principal Component Analysis (MPCA), the Tensor Rank-One Decomposition (TROD), Tensor-Train PCA (TT-PCA) and Hierarchical Tucker PCA (HT-PCA)  methods which utilize the models introduced in Section~\ref{sec:TensorDecomps} in order to learn a common subspace for a collection of tensors in a given training set. This common subspace is then used to project test tensor samples into lower-dimensional spaces and classify them \cite{Lu_Multilinear_2013,Lu_MPCA_2008}. This framework has found applications in supervised learning settings, in particular for face and object recognition images collected across different modalities and angles \cite{Vasilescu_Multilinear_2002,bengua2017matrix,wang2018principal,chaghazardi2017sample}.  Next, in Section~\ref{sec:RobustVersions}, we address the issue of robust low-rank tensor recovery for grossly corrupted and noisy higher order data for the different tensor models introduced in Section~\ref{sec:TensorDecomps}. Finally, in Section~\ref{sec:Conc} we provide an overview of ongoing work in the area of large tensor data factorization and computationally efficient implementation of the existing methods.

\section{Notation, Tensor Basics, and Preludes to Tensor PCA}
\label{sec:Setup}

Let $[n] := \{ 1, \dots, n \}$ for all $n \in \mathbb{N}$.  A \textit{$d$-mode}, or \textit{$d$th-order}, tensor is simply a $d$-dimensional array of complex valued data $\mathcal{A} \in \mathbb{C}^{n_1 \times n_2 \times \dots \times n_d}$ for given dimension sizes $n_1, n_2, \dots, n_d \in \mathbb{N}$.  Given this, each entry of a tensor is indexed by an index vector ${\bf i} = ({i_1,i_1,\dots,i_{d}}) \in [n_1] \times [n_2] \times \dots \times [n_{d}]$.  The entry of $\mathcal{A}$ indexed by ${\bf i}$ will always be denoted by $a({\bf i}) =  a(i_1,i_2,\dots,i_{d}) = a_{i_1,i_2,\dots,i_{d}} \in \mathbb{C}$.  The $j^{\rm th}$ entry position of the index vector ${\bf i}$ for a tensor $\mathcal{A}$ will always be referred to as the $j^{\rm th}$-mode of $\mathcal{A}$. In the remainder of this paper vectors are always bolded, matrices capitalized, tensors of order potentially $\geq 3$ italicized, and tensor entries/scalars written in lower case.

\subsection{Fibers, Slices, and Other Sub-tensors}

When encountered with a higher order tensor $\mathcal{A}$ it is often beneficial to look for correlations across its different modes.  For this reason, some of the many lower-order sub-tensors contained within any given higher-order tensor have been given special names and thereby elevated to special status.  In this subsection we will define a few of these.

\begin{figure}[h t]
 \centering
 \begin{subfigure}{0.2\textwidth}
  \centering
  \includegraphics[width=1\linewidth]{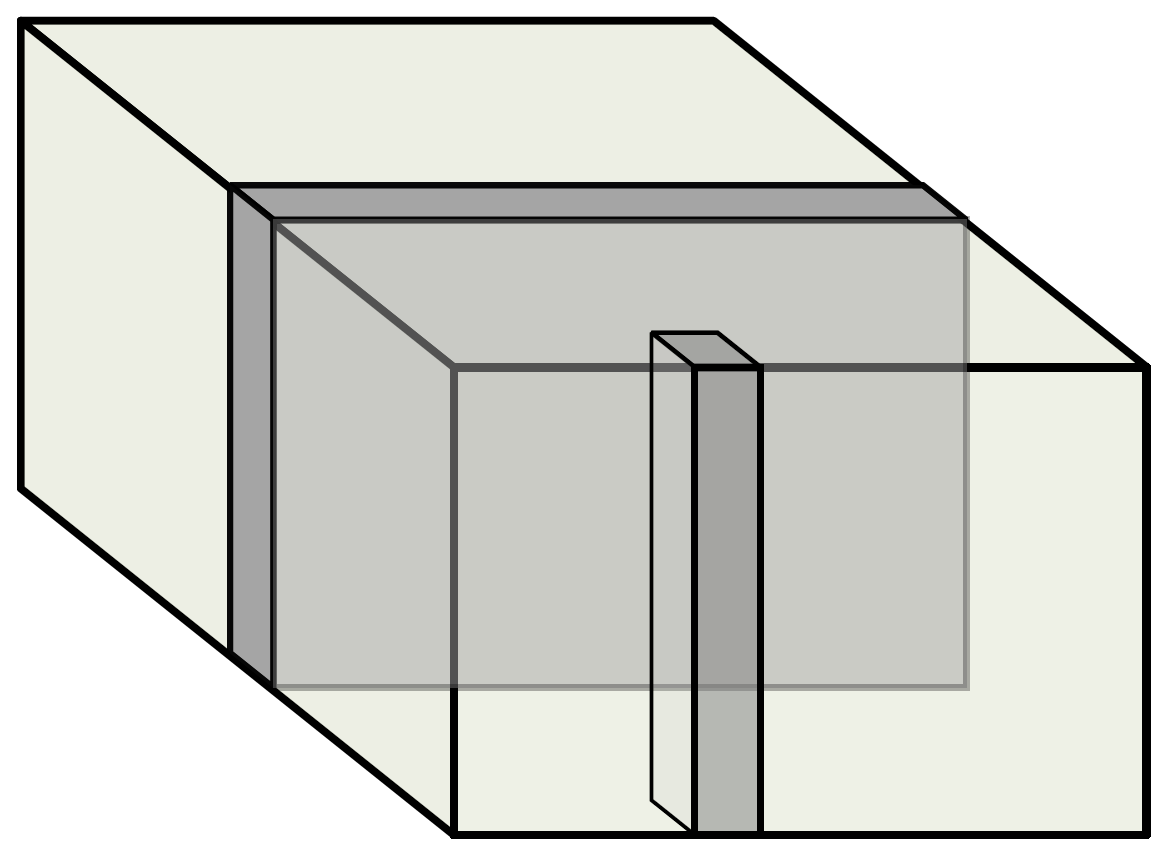}
  \caption{}
  \label{fig:tensor-a}
 \end{subfigure} \hspace{3mm}
 \begin{subfigure}{0.3\textwidth}
  \centering
  \includegraphics[width=1\linewidth]{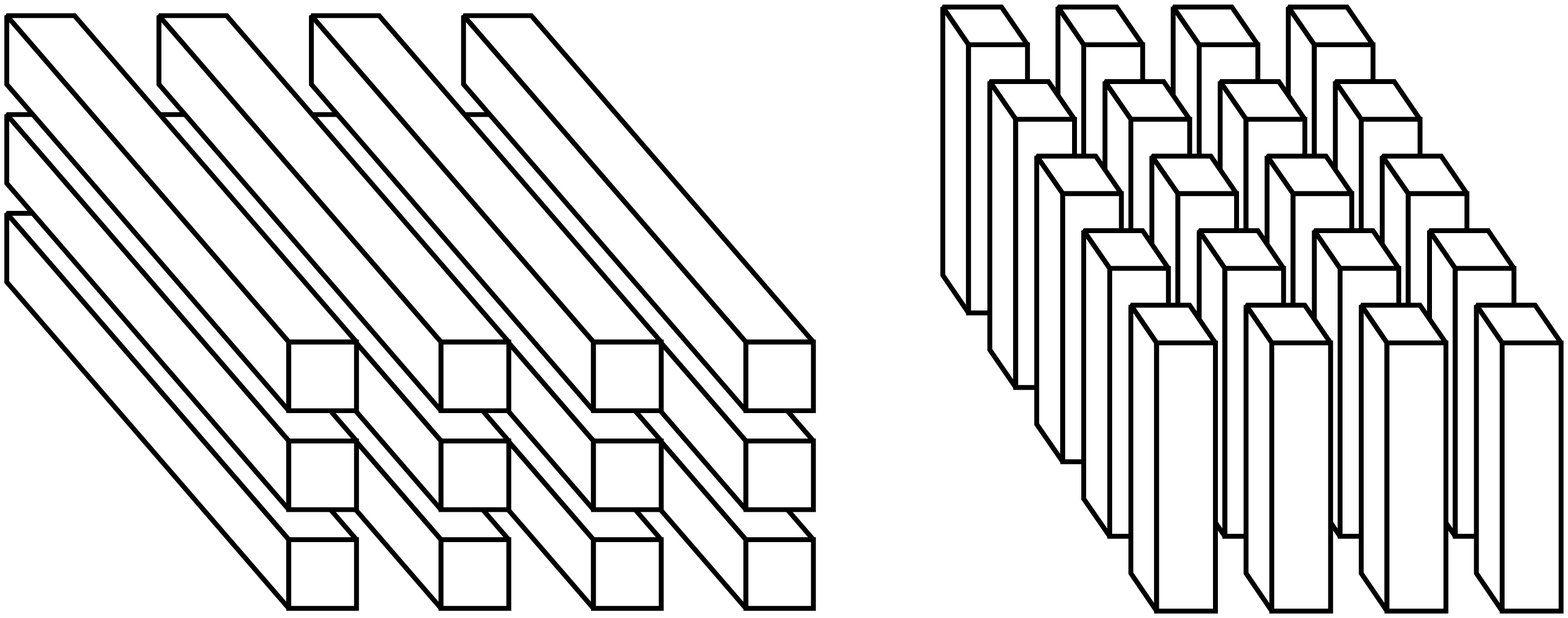}
  \caption{}
  \label{fig:tensor-b}
 \end{subfigure} \hspace{3mm}
 \begin{subfigure}{0.2\textwidth}
 \includegraphics[width=1\linewidth]{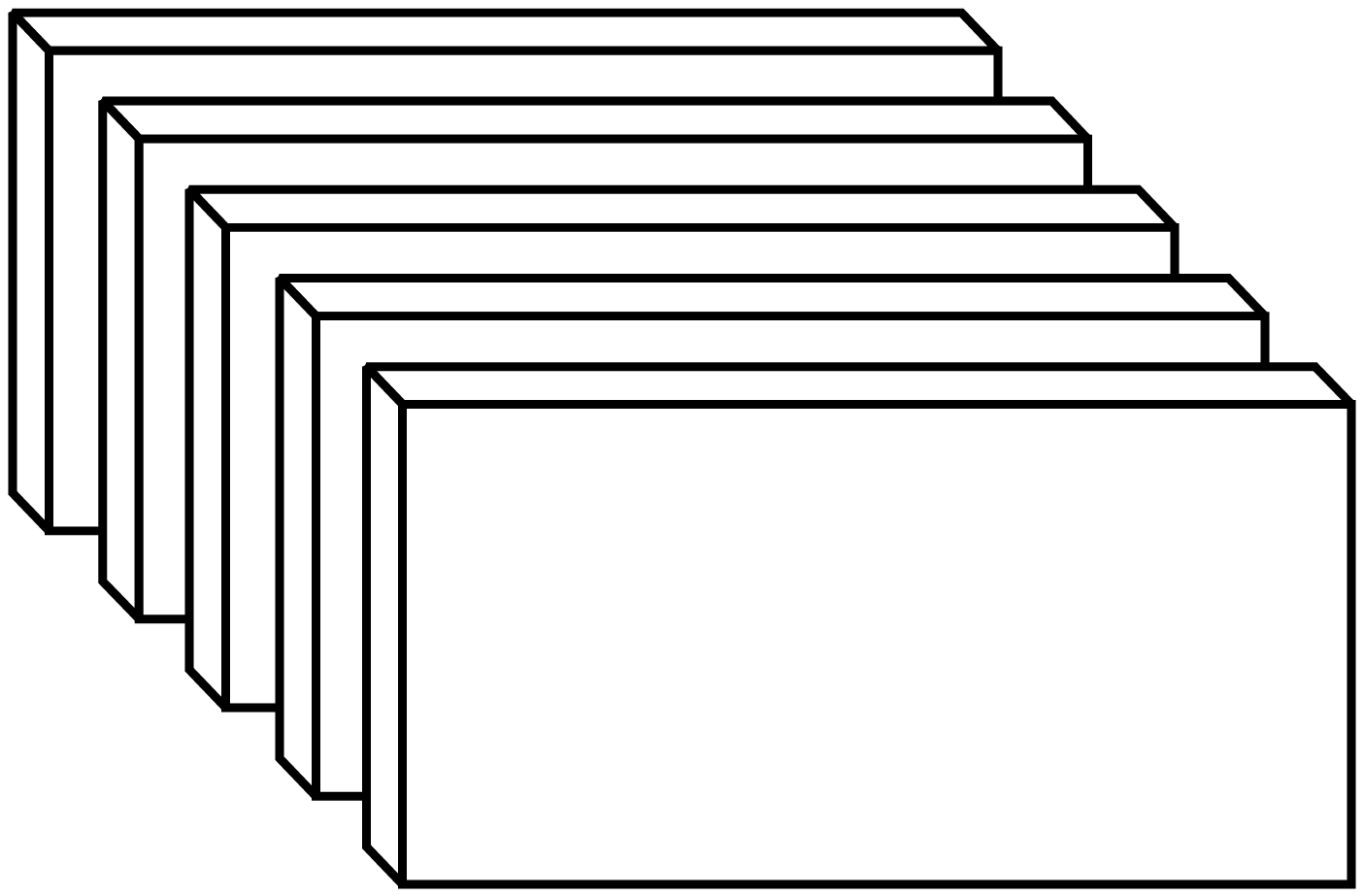}
  \caption{}
  \label{fig:tensor-c}
 \end{subfigure}

 \caption{Tensor, fibers and slices. (a) A $3$-mode tensor with a fiber and slice. (b) Left: mode-$3$ fibers. Right: mode-$1$ fibers. (c) mode-$3$ slices.}
 \label{fig:tensor}
\end{figure}

{\it Fibers} are $1$-mode sub-tensors (i.e., sub-vectors) of a given $d$-mode tensor $\mathcal{A} \in \mathbb{C}^{n_1 \times n_2 \times \dots \times n_d}$.  More specifically, a {\it mode-$j$ fiber} of $\mathcal{A}$ is a $1$-mode sub-tensor indexed by the $j^{\rm th}$ mode of $\mathcal{A}$ for any given choices of $j \in [d]$ and $i_\ell \in [n_\ell]$ for all $\ell \in [d] \setminus \{j\}$.   Each such mode-$j$ fiber is denoted by
\begin{equation}
{\bf a}(i_1, \dots, i_{j-1}, {\bf:}, i_{j+1}, \dots, i_d) = {\bf a}_{i_1, \dots, i_{j-1}, {\bf:}, i_{j+1}, \dots, i_d} \in \mathbb{C}^{n_j}.
\end{equation}
The $k^{\rm th}$ entry of a mode-$j$ fiber is $a(i_1, \dots, i_{j-1}, k, i_{j+1}, \dots, i_d) = a_{i_1, \dots, i_{j-1}, k, i_{j+1}, \dots, i_d} \in \mathbb{C}$ for each $k \in [n_j]$.  Note that there are $\prod_{k \in [d] \setminus \{ j \}} n_k$ mode-$j$ fibers of any given $\mathcal{A} \in \mathbb{C}^{n_1 \times n_2 \times \dots \times n_d}$.

\begin{examp}
Consider a $3$-mode tensor $\mathcal{A} \in \mathbb{C}^{m \times n \times p}$.  Its mode-$3$ fiber for any given $(i,j) \in [m] \times [n]$ is the $1$-mode sub-tensor ${\bf a}(i, j, {\bf:}) = {\bf a}_{i, j, {\bf:}} \in \mathbb{C}^p$.  There are $m n$ such mode-$3$ fibers of $\mathcal{A}$. Fibers of a $3$-mode tensor $\mathcal{A} \in \mathbb{C}^{3 \times 4 \times 5}$ are depicted in Figure \ref{fig:tensor}(b).
\end{examp}

In this paper {\it slices}, will always be $(d-1)$-mode sub-tensors of a given $d$-mode tensor $\mathcal{A} \in \mathbb{C}^{n_1 \times n_2 \times \dots \times n_d}$.  In particular, a {\it mode-$j$ slice} of $\mathcal{A}$ is a $(d-1)$-mode sub-tensor indexed by all but the $j^{\rm th}$ mode of $\mathcal{A}$ for any given choice of $j \in [d]$.  Each such mode-$j$ slice is denoted by
\begin{equation}
\mathcal{A}_{i_j = k} = \mathcal{A}({\bf:}, \dots, {\bf:}, k, {\bf:}, \dots, {\bf:}) = \mathcal{A}_{{\bf:}, \dots, {\bf:}, k, {\bf:}, \dots, {\bf:}} \in \mathbb{C}^{n_1 \times \dots \times n_{j-1} \times n_{j+1} \times \dots \times n_d}
\end{equation}
for any choice of $k \in [n_j]$.  The $(i_1, \dots, i_{j-1}, i_{j+1}, \dots, i_d)^{\rm th}$ entry of a mode-$j$ slice is
\begin{equation}
(a_{i_j=k}) ({i_1, \dots, i_{j-1}, i_{j+1}, \dots, i_{d}}) = a(i_1, \dots, i_{j-1}, k, i_{j+1}, \dots, i_d) = a_{i_1, \dots, i_{j-1}, k, i_{j+1}, \dots, i_d} \in \mathbb{C}
\end{equation}
for each $(i_1, \dots, i_{j-1}, i_{j+1}, \dots, i_d) \in [n_1] \times \dots \times [n_{j-1}] \times [n_{j+1}] \times \dots \times [n_d]$.  It is easy to see that there are always just $n_j$ mode-$j$ slices of any given $\mathcal{A} \in \mathbb{C}^{n_1 \times n_2 \times \dots \times n_d}$.

\begin{examp}
Consider a $3$-mode tensor $\mathcal{A} \in \mathbb{C}^{m \times n \times p}$.  Its mode-$3$ slice for any given $k \in [p]$ is the $2$-mode sub-tensor (i.e., matrix) $A_{i_3 = k} = A({\bf:}, {\bf:}, k) = A_{{\bf:}, {\bf:}, k} \in \mathbb{C}^{m \times n}$.  There are $p$ such mode-$3$ slices of $\mathcal{A}$.  In Figure \ref{fig:tensor}(c), the $5$ mode-$3$ slices of $\mathcal{A} \in \mathbb{C}^{3 \times 4 \times 5}$ can be viewed.
\end{examp}

\subsection{Tensor Vectorization, Flattening, and Reshaping}

There are a tremendous multitude of ways one can reshape a $d$-mode tensor into another tensor with a different number of modes.  Perhaps most important among these are the transformation of a given tensor $\mathcal{A} \in \mathbb{C}^{n_1 \times n_2 \times \dots \times n_d}$ into a vector or matrix so that methods from standard numerical linear algebra can be applied to the reshaped tensor data thereafter.

The {\it vectorization} of $\mathcal{A} \in \mathbb{C}^{n_1 \times n_2 \times \dots \times n_d}$ will always reshape $\mathcal{A}$ into a vector (i.e., $1$st-order tensor) denoted by ${\bf a} \in \mathbb{C}^{n_1 n_2 \cdots n_d}$.  This process can be accomplished numerically by, e.g., recursively vectorizing the last two modes of $\mathcal{A}$ (i.e., each matrix $A(i_1, \dots, i_{d-2}, {\bf:}, {\bf:}))$ according to their row-major order until only one mode remains.  When done in this fashion the entries of the vectorization ${\bf a}$ can be rapidly retrieved from $\mathcal{A}$ via the formula
\begin{equation}
a_{j} = \mathcal{A}(g_1(j), \dots, g_d(j)),
\end{equation}
where each of the index functions $g_m: [n_1 n_2 \cdots n_d] \rightarrow [n_m]$ is defined for all $m \in [d]$ by
\begin{equation}
g_{m}(j) := \left \lceil \frac{j}{\prod_{\ell \in [d]\setminus [m] } n_\ell} \right \rceil ~{\rm mod}~ n_{m} + 1.
\end{equation}
Herein we will always use the convention that $\prod_{\ell \in \emptyset} n_\ell := 1$ to handle the case where, e.g., $[d]\setminus [m]$ is the empty set above.

The process of reshaping a $(d > 2)$-mode tensor into a matrix is known as {\it matricizing}, {\it flattening}, or {\it unfolding}, the tensor.  There are $2^d - 2$ nontrivial ways in which one may create a matrix from a $d$-mode tensor by partitioning its $d$ modes into two different ``row'' and ``column'' subsets of modes (each of which is then implicitly vectorized separately).\footnote{One can use Stirling numbers of the second kind to easily enumerate all possible mode partitions.}  The most often considered of these are the mode-$j$ variants mentioned below (excluding, perhaps, the alternate matricizations utilized as part of, e.g., the tensor train \cite{Oseledets_Tensor_2011} and hierarchical SVD \cite{Grasedyck_Hierarchical_2010} decomposition methods.)

\begin{figure}[H]
\centering
\includegraphics[scale=0.2]{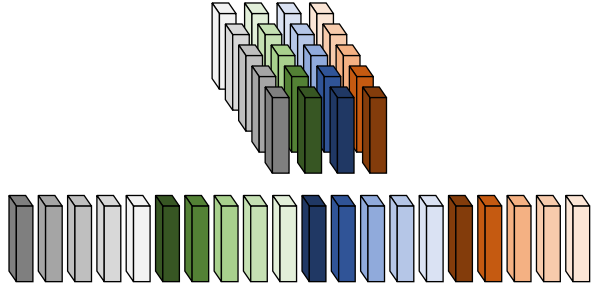}
\caption{Formation of the mode-$1$ unfolding (below) using the mode-$1$ fibers of a $3$-mode tensor (above).}
\label{fig:unfolding}
\end{figure}

The \textit{mode-$j$ matricization}, \textit{mode-$j$ flattening}, or \textit{mode-$j$ unfolding} of a tensor $\mathcal{A} \in \mathbb{C}^{n_1 \times n_2 \times \dots \times n_d}$, denoted by $A^{(j)} \in \mathbb{C}^{n_j \times \prod_{\ell \in [d] \setminus \{ j \}} n_\ell}$, is a matrix whose columns consist of all the mode-$j$ fibers of $\mathcal{A}$.  More explicitly, $A^{(j)}$ is determined herein by defining its entries to be
\begin{equation}
\left( a^{(j)} \right)_{k,l} := \mathcal{A}\left(h_1 (l),h_2(l),\dots, h_{j-1}(l), k ,h_{j+1}(l), \dots, h_{d-1}(l), h_{d}(l) \right),
\end{equation}
where, e.g., the index functions $h_m: \left[ \prod_{\ell \in [d] \setminus \{ j \}} n_\ell \right] \rightarrow [n_m]$ are defined by
\begin{equation}
h_{m}(l) := \left \lceil \frac{l}{\prod_{\ell \in [d]\setminus \left( [m] \cup \{ j \} \right) } n_\ell} \right \rceil ~{\rm mod}~ n_{m} +1
\end{equation}
for all $m \in [d] \setminus \{j \}$ and $l \in \prod_{\ell \in [d] \setminus \{ j \}} n_\ell$.
Note that $A^{(j)}$'s columns are ordered by varying the index of the largest non-$j$ mode ($d$ unless a mode-$d$ unfolding is being constructed) fastest, followed by varying the second largest non-$j$ mode second fastest, etc..  %%Though an arbitrary convention, this particular choice for organizing the columns of $A^{(j)}$ will make some notation less cumbersome later.
In Figure \ref{fig:unfolding}, the mode-$1$ matricization of $\mathcal{A} \in \mathbb{C}^{3 \times 4 \times 5}$ is formed in this way using its mode-$1$ fibers.

\begin{examp}
As an example, consider a $3$-mode tensor $\mathcal{A}\in \mathbb{C}^{n_1\times n_2\times n_3}$.  Its mode-$1$ matricization is
\begin{equation}
A^{(1)} = \left[ \left(A_{i_2 = 1} \right)^{(1)} \Big|  \left( A_{i_2 = 2} \right)^{(1)} \Big| \dots \Big| \left( A_{i_2 = n_2 } \right)^{(1)} \right] \in \mathbb{C}^{n_1 \times n_2 n_3}
\label{equ:Matricizeexamp1}
\end{equation}
where $\left(A_{i_2 = c} \right)^{(1)} \in \mathbb{C}^{n_1 \times n_3}$ is the mode-$1$ matricization of the mode-$2$ slice of $\mathcal{A}$ at $i_2 = c$ for each $c \in [n_2]$.  Note that we still consider each mode-$2$ slice $A_{i_2 = c}$ to be indexed by $i_1$ and $i_3$ above.  As a result, e.g., $\left(A_{i_2 = 1} \right)^{(3)}$ {\it will be considered meaningful} while $\left(A_{i_2 = 1} \right)^{(2)}$ {\it will be considered meaningless} in the same way that, e.g., $\left(A_{i_2 = 1} \right)^{(100)}$ is meaningless.  That is, even though $A_{i_2 = c}$ only has two modes, we will still consider them to be its ``first'' and ``third'' modes (i.e., its two mode positions are still indexed by $1$ and $3$, respectively).  Though potentially counterintuitive at first, this notational convention will simplify the interpretation of expressions like \eqref{equ:Matricizeexamp1} -- \eqref{equ:Matricizeexamp3} going forward.

Given this convention, we also have that
\begin{equation}
A^{(2)} = \left[ \left(A_{i_1 = 1} \right)^{(2)} \Big|  \left( A_{i_1 = 2} \right)^{(2)} \Big| \dots \Big| \left( A_{i_1 = n_1} \right)^{(2)} \right] \in \mathbb{C}^{n_2 \times n_1 n_3},
\end{equation}
and
\begin{equation}
A^{(3)} = \left[ \left(A_{i_1 = 1} \right)^{(3)} \Big|  \left( A_{i_1 = 2} \right)^{(3)} \Big| \dots \Big| \left( A_{i_1 = n_1} \right)^{(3)} \right] \in \mathbb{C}^{n_3 \times n_1 n_2}.
\label{equ:Matricizeexamp3}
\end{equation}
\end{examp}

\subsection{The Standard Inner Product Space of $d$-mode Tensors}

The set of all $d$th-order tensors $\mathcal{A} \in \mathbb{C}^{n_1 \times n_2 \times \dots \times n_d}$ forms a vector space over the complex numbers when equipped with componentwise addition and scalar multiplication.  This vector space is usually endowed with the Euclidean inner product.  More specifically, the inner product of $\mathcal{A}, \mathcal{B} \in \mathbb{C}^{n_1\times n_2\times \ldots \times n_{d}}$ will always be given by

\begin{equation}
\left\langle \mathcal{A},\mathcal{B} \right\rangle := {\sum_{i_1=1}^{n_1} \sum_{i_2=1}^{n_2} \ldots\sum_{i_{d}=1}^{n_{d}} a(i_1,i_2,\ldots,i_{d}) \overline{b(i_1,i_2,\ldots,i_{d})}},
\end{equation}
where $\overline{(\cdot)}$ denotes the complex conjugation operation. This inner product then gives rise to the standard Euclidean norm

\begin{equation}
\| \mathcal{A} \| := \sqrt{\left\langle \mathcal{A},\mathcal{A} \right\rangle} = \sqrt{\sum_{i_1=1}^{n_1} \sum_{i_2=1}^{n_2} \ldots \sum_{i_{d}=1}^{n_{d}} \left| a(i_1,i_2,\ldots,i_{d}) \right|^2}.
\end{equation}
If $\left\langle \mathcal{A},\mathcal{B} \right\rangle = 0$ we will say that $\mathcal{A}$ and $\mathcal{B}$ are {\it orthogonal}.  If $\mathcal{A}$ and $\mathcal{B}$ are orthogonal and also have unit norm (i.e., have $\| \mathcal{A} \| = \| \mathcal{B} \| = 1$) we will say that they are {\it orthonormal}.

It is worth noting that trivial inner product preserving isomorphisms exist between this standard inner product space and any of its reshaped versions (i.e., reshaping can be viewed as an isomorphism between the original $d$-mode tensor vector space and its reshaped target vector space).  In particular, the process of reshaping tensors is linear.  If, for example, $\mathcal{A}, \mathcal{B} \in \mathbb{C}^{n_1\times n_2\times \ldots \times n_{d}}$ then one can see that the mode-$j$ unfolding of $\mathcal{A} + \mathcal{B} \in \mathbb{C}^{n_1\times n_2\times \ldots \times n_{d}}$ is $\left( \mathcal{A} + \mathcal{B} \right)^{(j)} = A^{(j)} + B^{(j)}$ for all $j \in [d]$.  Similarly, the vectorization of $\mathcal{A} + \mathcal{B}$ is always exactly ${\bf a} + {\bf b}$.

\subsection{Tensor Products and $j$-Mode Products}

It is occasionally desirable to build one's own higher order tensor using two lower order tensors.  This is particularly true when one builds them up using vectors as part of, e.g., PARAFAC/CANDECOMP decomposition techniques \cite{Harshman_PARAFAC_1970,Kruskal_Three_1977,Bro_PARAFAC_1997,Faber_Recent_2003,Kolda_Tensor_2009}.  Toward this end we will utilize the \textit{tensor product} of two tensors $\mathcal{A} \in \mathbb{C}^{n_1 \times n_2 \times \dots \times n_d}$ and $\mathcal{B} \in \mathbb{C}^{n'_1 \times n'_2 \times \dots \times n'_{d'}}$.  The result of the tensor product, $\mathcal{A} \otimes \mathcal{B} \in \mathbb{C}^{n_1 \times n_2 \times \dots \times n_d \times n'_1 \times n'_2 \times \dots \times n'_{d'}}$, is a $(d + d')$-mode tensor whose entries are given by
\begin{equation}
\left( \mathcal{A} \otimes \mathcal{B} \right)_{i_1, \dots, i_d, i'_1, \dots, i'_{d'}} = a(i_1, \dots, i_d) \overline{b(i'_1, \dots, i'_{d'})}.
\end{equation}

A $d$th-order tensor which is built up from $d$ vectors using the tensor product is called a {\it rank-$1$ tensor}.  For example, $\bigotimes^4_{k=1} {\bf a}_k = {\bf a}_1 \otimes {\bf a}_2 \otimes {\bf a}_3 \otimes {\bf a}_4 \in \mathbb{C}^{n_1 \times n_2 \times n_3 \times n_4}$ is a rank-$1$ tensor with $4$ modes which is built from ${\bf a}_1 \in \mathbb{C}^{n_1}$, ${\bf a}_2 \in \mathbb{C}^{n_2}$, ${\bf a}_3 \in \mathbb{C}^{n_3}$, and ${\bf a}_4 \in \mathbb{C}^{n_4}$.  Note that this $4$th-order tensor is unambiguously called ``rank-1'' due to the fact that ${\bf a}_1 \otimes {\bf a}_2 \otimes {\bf a}_3 \otimes {\bf a}_4 \in \mathbb{C}^{n_1 \times n_2 \times n_3 \times n_4}$ is both built up from rank-$1$ tensors, {\it and} because every mode-$j$ unfolding of ${\bf a}_1 \otimes {\bf a}_2 \otimes {\bf a}_3 \otimes {\bf a}_4$ is also a rank-$1$ matrix.

Finally, the \textit{mode-$j$ product} of $d$-mode tensor $\mathcal{A} \in \mathbb{C}^{n_1 \times \dots \times n_{j-1} \times n_j \times n_{j+1} \times \dots \times n_d}$ with a matrix $U \in \mathbb{C}^{m_j \times n_j}$ is another $d$-mode tensor $\mathcal{A} \times_j U \in \mathbb{C}^{n_1 \times \dots \times n_{j-1} \times m_j \times n_{j+1} \times \dots \times n_d}$.  Its entries are given by
\begin{equation}
(\mathcal{A} \times_j U)_{i_1, \dots, i_{j-1}, \ell, i_{j+1}, \dots, i_d} = \sum_{i_{j}=1}^{n_j} a_{i_{1},\dots,i_j,\dots,i_{d}} u_{\ell,i_{j}}
\label{equ:DefModejProduct}
\end{equation}
for all $(i_1, \dots, i_{j-1}, \ell, i_{j+1}, \dots, i_d) \in [n_1] \times \dots \times [n_{j-1}] \times [m_j] \times [n_{j+1}] \times \dots \times [n_d]$.  Looking at the mode-$j$ unfoldings of $\mathcal{A} \times_j U$ and $\mathcal{A}$ one can easily see that $(\mathcal{A} \times_j U)^{(j)} = U A^{(j)}$ holds for all $j \in [d]$.

Mode-$j$ products play a particularly important role in many tensor PCA and tensor factorization methods \cite{Tucker_Extension_1964,Lathauwer_Multilinear_2000,Vasilescu_Multilinear_2002,Kolda_Tensor_2009}.  For this reason it is worth stating some of their basic properties:  mainly, mode-$j$ products are bilinear, commute on different modes, and combine in reverse order on the same mode.  The  following simple lemma formally lists these important properties.

\begin{lem}
Let $\mathcal{A}, \mathcal{B} \in \mathbb{C}^{n_1 \times n_2 \times \dots \times n_d}$, $\alpha, \beta \in \mathbb{C}$, and $U_\ell, V_\ell \in \mathbb{C}^{m_\ell \times n_\ell}$ for all $\ell \in [d]$.  The following four properties hold:
\begin{enumerate}
\item[($\dagger$)] $\left( \alpha \mathcal{A} + \beta \mathcal{B} \right) \times_j U_j = \alpha \left( \mathcal{A} \times_j U_j \right) + \beta \left( \mathcal{B} \times_j U_j \right)$.\\

\item[($\dagger \dagger$)] $\mathcal{A} \times_j  \left( \alpha U_j + \beta V_j \right) = \alpha \left( \mathcal{A} \times_j U_j \right) + \beta \left( \mathcal{A} \times_j V_j \right)$. \\

\item[($\dagger \dagger \dagger$)] If $j \neq \ell$ then $\mathcal{A} \times_j U_j \times_\ell V_\ell = \left( \mathcal{A} \times_j U_j \right) \times_\ell V_\ell = \left( \mathcal{A} \times_\ell V_\ell \right) \times_j U_j = \mathcal{A} \times_\ell V_\ell \times_j U_j$ .\\

\item[($\dagger \dagger \dagger \dagger$)] If $W \in \mathbb{C}^{p \times m_j}$ then $\mathcal{A} \times_j U_j \times_j W = \left( \mathcal{A} \times_j U_j \right) \times_j W = \mathcal{A} \times_j \left( WU_j \right) = \mathcal{A} \times_j WU_j $.
\end{enumerate}

\label{lem:modeProdProps}
\end{lem}

Another way to represent tensor products is to define index contractions. An \textit{index contraction} is the sum over all the possible values of the repeated indices of a set of tensors. For example, the matrix product $C_{\alpha\gamma}=\sum_{\beta=1}^{D}A_{\alpha\beta}B_{\beta\gamma}$ can be thought of the contraction of index $\beta$. One can also consider tensor products through index contractions such as $F_{\gamma\omega\rho\sigma}=\sum_{\alpha,\beta,\delta,\nu,\mu=1}^{D}A_{\alpha\beta\delta\sigma}B_{\beta\gamma\mu}C_{\delta\nu\mu\omega}E_{\nu\rho\alpha}$, where the indices $\alpha,\beta,\delta,\nu,\mu$ are contracted to produce a new four-mode tensor. The representation of a tensor through the contraction of indices of other tensors will be particularly important for the study of tensor networks (TNs).

%%%%%%%%%%%%%%%%%%%%%%%%%%%%%%%%%%%%%%%%%%%
% TENSOR FACTORIZTION AND COMPRESSION
%%%%%%%%%%%%%%%%%%%%%%%%%%%%%%%%%%%%%%%%%%%

\section{Tensor Factorization and Compression Methods}
\label{sec:TensorDecomps}
In this section, we focus on tensor decomposition methods that provide low-rank approximations for multilinear datasets by reducing their complexity similar to the way PCA/SVD does for matrices.  Advantages of using multiway analysis over two-way analysis in terms of uniqueness, robustness to noise, and computational complexity have been shown in many studies (see, e.g., \cite{acar2009unsupervised, acar2005modeling, estienne2001multi}). In this section we review some of the most commonly used tensor models for representation and compression and present results on their uniqueness and storage complexities.  We then empirically evaluate the compression versus approximation error performance of several of these methods for three different higher order datasets.

\subsection{CANDECOMP/PARAFAC Decomposition (CPD)}
CPD is a generalization of PCA to higher order array and represents a $d$-mode tensor $\mathcal{X} \in \mathbb{R}^{n_1 \times n_2 \times ...\; \times n_d}$ as a combination of rank-one tensors  \cite{kruskal1977three}.
\begin{equation}
\mathcal{X} = \sum_{r=1}^R \lambda_r {\bf a}_r^{(1)} \otimes {\bf a}_r^{(2)} \otimes .... \otimes {\bf a}_r^{(d)},
\end{equation}
\noindent where $R$ is a positive integer, $\lambda_r$ is the weight of the $r$th rank-one tensor, ${\bf a}_r^{(i)}\in \mathbb{R}^{n_i}$ is the $r$th factor of $i$th mode with unit norm where $i\in [d]$ and $r\in [R]$. Alternatively, $\mathcal{X}$ can be represented as mode products of a diagonal core tensor $\mathcal{S}$ with entries $s(i,i,...,i)=\lambda_i$ and factor matrices $ A^{(i)}=[{\bf a}_1^{(i)} {\bf a}_2^{(i)} ...\; {\bf a}_R^{(i)}]$ for $i \in [d]$:
\begin{equation}
\mathcal{X}= \mathcal{S} \times_1 { A}^{(1)} \times_2 {A}^{(2)} ... \times_d {A}^{(d)}.
\end{equation}
The main restriction of the PARAFAC model is that the factors across different modes only interact factorwise. For example, for a $3$-mode tensor, the $i$th factor corresponding to the first mode only interacts with the $i$th factors of the second and third modes.

Rank-R approximation of a $d$th-order tensor  $\mathcal{X}\in \mathbb{R}^{n_{1}\times n_{2}...\times n_{d}}$ with $n_{1}=n_{2}=\ldots=n_{d}=n$ obtained by CPD is represented using $\mathcal{O}(Rdn)$ parameters, which is less than the number of parameters required for PCA applied to an unfolded matrix.

\textit{Uniqueness}: In a recent review article, Sidiropoulos et al. \cite{sidiropoulos2017tensor} provided two alternative proofs for the uniqueness of the PARAFAC model. Given a tensor $\mathcal{X}$ of rank $R$, its PARAFAC decomposition is essentially unique, i.e. the factor matrices $A^{(1)},\ldots,A^{(d)}$ are unique up to a common permutation and scaling of columns for the given number of terms. Alternatively, Kruskal provided results  on uniqueness of $3$-mode CPD depending on matrix $k$-rank as:
\begin{equation}
k_{{ A}^{(1)}}+k_{{ A}^{(2)}}+k_{{ A}^{(3)}}\geq 2R+2,
\end{equation}
\noindent where $k_{{ A}^{(i)}}$ is the maximum value $k$ such that any $k$ columns of ${{ A}^{(i)}}$ are linearly independent \cite{kruskal1977three}. This result is later generalized for $d$-mode tensors in \cite{sidiropoulos2000uniqueness} as:
\begin{equation}
\sum_{i=1}^d k_{{ A}^{(i)}}\geq 2R+d-1.
\end{equation}
Under these conditions, the CPD solution is unique and the estimated model cannot be rotated without a loss of fit.

\textit{Computational Issues}: CPD is most commonly computed by alternating least squares (ALS) by successively assuming the factors in $d-1$ modes known and then estimating the unknown set of parameters of the last mode. For each mode and each iteration,  the Frobenious norm of the difference between input tensor and CPD approximation is minimized. ALS is an attractive method since it ensures the improvement of the solution in every iteration. However, in practice, the existence of large amount of noise or the high order of the model may prevent ALS to converge to global minima or require several thousands of iterations \cite{Kolda_Tensor_2009,cichocki2015tensor,  grasedyck2013literature}. Different methods have been proposed to improve performance and accelerate convergence rate of
CPD algorithms \cite{phan2013fast,chen2017h}. A number of particular techniques exist, such as line search extrapolation methods \cite{andersson2000n,rajih2008enhanced,chen2011new} and compression \cite{kiers1998three}. Instead of alternating
estimation, all-at-once algorithms such as the OPT algorithm \cite{acar2011scalable}, the conjugate gradient algorithm for nonnegative CP \cite{cohen2015fast},
the PMF3, damped Gauss-Newton (dGN) algorithms \cite{paatero1997weighted}
and fast dGN \cite{phan2013low} have been studied to deal with
problems of a slow convergence of the ALS in some cases.
Another approach is to consider the CP decomposition as a
joint diagonalization problem \cite{de2006link,de2008blind}.

\subsection{Tucker Decomposition and HoSVD} Tucker decomposition is a natural extension of the SVD to $d$-mode tensors and decomposes the tensor into a core tensor multiplied by a matrix along each mode \cite{de2000multilinear,cichocki2015tensor, grasedyck2013literature}. Tucker decomposition of a $d$-mode tensor $\mathcal{X} \in \mathbb{R}^{n_1 \times n_2 \times ...\; \times n_d}$ is written as:
\begin{equation}
\begin{array}{c}
\mathcal{X} = \sum_{i_1=1}^{n_1}...\sum_{i_d=1}^{n_d} s_{i_1,i_2,...,i_d} \left({\bf u}^{(1)}_{i_1}\otimes{\bf u}^{(2)}_{i_2}\otimes ...\otimes{\bf u}^{(d)}_{i_d}\right),\\
\mathcal{X} = \mathcal{S}\times_1 U^{(1)} \times_2 U^{(2)}...\times_d U^{(d)},
\label{eq:tucker}
\end{array}
\end{equation}
\noindent where the matrices $U^{(i)}=[{\bf u}_1^{(i)} {\bf u}_2^{(i)} ...\; {\bf u}_{n_d}^{(i)}]$s are square factor matrices and the core tensor $\mathcal{S}$ is obtained by $\mathcal{S} = \mathcal{X}\times_1 U^{(1),\top} \times_2 U^{(2),\top}...\times_d U^{(d),\top}$, where $U^{(i),\top}$ denotes the transpose of the factor matrix along each mode.  It is common for the Tucker decomposition to assume the rank of $U^{(i)}$s to be less than $n_{i}$ so that $\mathcal{S}$ is a compression of $\mathcal{X}$. Multilinear-rank-R approximation of a $d$th-order tensor  $\mathcal{X}\in \mathbb{R}^{n_{1}\times n...\times n_{d}}$ with $n_{1}=n_{2}=\ldots=n_{d}=n$ is represented using $\mathcal{O}(Rnd+R^d)$ parameters in the Tucker model.

In contrast to PARAFAC, Tucker models allow interactions between the factors obtained across the modes and  the core tensor includes the strength of these interactions. In summary, both CPD and Tucker are sum-of-outer products
models, and one can argue that the most general
form of one contains the other. However, what distinguishes the two
is uniqueness.

\textit{Uniqueness:} As it can be seen in (\ref{eq:tucker}), one can linearly transform the columns of $U^{(i)}$ and
absorb the inverse transformation in the core tensor $\mathcal{S}$.  The subspaces
defined by the factor matrices in Tucker decomposition are unique, while the bases
in these subspaces may be chosen arbitrarily -- their choice is
compensated for within the core tensor. For this reason, the Tucker model is
not unique unless additional constraints are placed on $U^{(i)}$s and/or the core
tensor. Constraints such as orthogonality, nonnegativeness, sparsity, independence and smoothness have been imposed on the factor matrices to obtain unique decompositions  \cite{cichocki2009nonnegative,zhou2012fast,zubair2013tensor,morup2008algorithms}.

The Higher Order SVD (HoSVD) is a special case of Tucker decomposition obtained by adding an orthogonality constraint to the component matrices. In HoSVD, the factor matrices, $U^{(i)}$s, are the left singular vectors of each flattening $ X^{(i)}$. In HoSVD, low n-rank approximation of $\mathcal{X}$ can be obtained by truncating the orthogonal factor matrices of HoSVD resulting in truncated HoSVD. Due to the orthogonality of the core tensor, HoSVD is unique for a specific multilinear rank.

As opposed to the SVD for matrices, the $(R_{1},R_{2},\ldots,R_{d})$ truncation of the HoSVD is not the best $(R_1,R_2,\ldots,R_d)$ approximation of $\mathcal{X}$.  The best $(R_1,R_2,\ldots,R_d)-$rank approximation is obtained by solving the following optimization problem.

\begin{equation}
\begin{array}{c}
\min_{\mathcal{S},U^{(1)},U^{(2)},\ldots,U^{(d)}}\left\| \mathcal{X} - \mathcal{S}\times_1 U^{(1)} \times_2 U^{(2)}\ldots \times_d U^{(d)} \right\| \\
\mbox{subject to}\\ \mathcal{S}\in \mathbb{R}^{R_1\times R_2\times \ldots \times R_d}, \\ U^{(i)}\in \mathbb{R}^{n_i\times R_i} \mbox{and columnwise orthogonal for all}\; i\in[d].
\end{array}
\end{equation}

\textit{Computational Issues:} It has been shown that this optimization problem can be solved by the ALS approach  iteratively and the method is known as higher-order orthogonal iteration (HOOI) \cite{de2000multilinear}. For many applications, HoSVD is considered to be sufficiently good, or it can serve as an initial value in algorithms for finding the best approximation \cite{bergqvist2010higher}.

To identify hidden nonnegative patterns in a tensor, nonnegative matrix factorization algorithms have been adapted to Tucker model \cite{morup2008algorithms, kim2007nonnegative,kim2008nonnegative,
	cichocki2007non, cichocki2009nonnegative}. NTD of a tensor $\mathcal{X} \in \mathbb{R}^{n_1 \times n_2 \times ...\; \times n_d}$ can be obtained by solving:

\begin{equation}
\begin{array}{c}
\min_{\mathcal{S},U^{(1)},U^{(2)},\ldots,U^{(d)}}\left \| \mathcal{X} - \mathcal{S}\times_1 U^{(1)} \times_2 U^{(2)}\ldots\times_d U^{(d)}\right \| \\
\mbox{subject to}\\ \mathcal{S}\in \mathbb{R}_+^{R_1\times R_2\times \ldots \times R_d}, \;\; U^{(i)}\in \mathbb{R}_+^{n_i\times R_i}; i\in[d].
\end{array}
\end{equation}

This optimization problem can be solved using nonnegative ALS and updating the core tensor $\mathcal S$ and factor matrices ${\bf U}^{(i)}$ at each iteration depending on different updating rules such as alpha and beta divergences \cite{cichocki2007non, kim2008nonnegative} or low-rank NMF \cite{morup2008algorithms, zhou2012fast}.
\subsection{Tensor Networks}
Tensor decompositions such as PARAFAC and Tucker decompose complex high dimensional data tensors into their factor tensors and matrices. Tensor networks (TNs), on the other hand, represent a higher-order tensor as a set of sparsely interconnected lower-order tensors, typically 3rd-order and 4th-order tensors called core and provide computational and storage benefits \cite{orus2014practical,cichocki2014era,cichocki2016tensor}. More formally, a TN is a set of tensors where some, or all, of its indices are contracted according to some pattern. The contraction of a TN with some open indices results in another tensor. One important property of TN is that the total number of operations that must be done to obtain the final result of a TN contraction depends on the order in which indices in the TN are contracted, i.e., for a given tensor there are many different TN representations and finding the optimal order of indices to be contracted is a crucial step in the efficiency of TN decomposition. The optimized topologies yield simplified and convenient graphical representations of higher-order tensor data \cite{handschuh2012changing,hubener2010concatenated}. Some commonly encountered tensor network topologies include Hierarchical Tucker (HT), Tree Tensor Network State (TTNS), Tensor Train (TT) and tensor networks with cycles such as Projected Entangled Pair States (PEPS) and Projected Entangled Pair Operators (PEPO).\\
\textit{ Uniqueness:} As noted above, for a given tensor there are many different TN representations, so in general, there is not a unique representation. The uniqueness of various TN models under different constraints is still an active field of research \cite{cichocki2016tensor}.
\subsubsection{Hierarchical Tensor Decomposition}

To reduce the memory requirements of Tucker decomposition,  hierarchical Tucker (HT) decomposition (also called hierarchical tensor representation) has been proposed \cite{hackbusch2009new,Grasedyck_Hierarchical_2010, grasedyck2013literature}.
Hierarchical Tucker Decomposition recursively splits the modes based on a hierarchy and creates a binary tree $T$ containing a subset of the modes $t \subset [d]$ at each node \cite{Grasedyck_Hierarchical_2010}. Factor matrices ${U}_t$s are obtained from the SVD of ${X}^{(t)}$ which is the matricization of a tensor $\mathcal{X}$ corresponding to the subset of the modes $t$ at each node. However, this matricization is different from mode-n matricization of the tensor, and rows of ${X}^{(t)}$ correspond to the modes in the set of $t$ while columns of ${X}^{(t)}$ store indices of the remaining modes. Constructed tree structure yields hierarchy amongst the factor matrices ${U}_t$ whose columns span ${X}^{(t)}$ for each $t$. Let $t_1$ and $t_2$ be children of $t$. For $t = t_1 \cup  t_2$ and $t_1 \cap t_2= \emptyset$, there exists a transfer matrix ${B}_t$ such that ${U}_t=({U}_{t_1} \otimes {U}_{t_2}){B}_t$, where ${B}_t\in \mathbb{R}^{R_{t_1}R_{t_2}\times R_t}$. By assuming $R_t=R$,  HT-rank-R approximation of $\mathcal{X}\in \mathbb{R}^{n_{1}\times n_{2}\times \ldots \times n_{d}}$ with $n_{1}=n_{2}=\ldots=n_{d}=n$ requires storing ${U}_t$s for the leaf nodes and ${B}_t$s for the other nodes in $T$ with  $\mathcal{O}(dnR + dR^3)$ parameters \cite{grasedyck2013literature}.

\subsubsection{Tensor Train Decomposition}
Tensor Train (TT) Decomposition can be interpreted as a special case of the HT, where all nodes of the underlying tensor network are connected in a cascade or train. It has been proposed to compress large tensor data into smaller core tensors \cite{Oseledets_Tensor_2011}. This model allows users to avoid the exponential growth of Tucker model  and provides more efficient storage complexity. TT decomposition of a tensor $\mathcal{X} \in \mathbb{R}^{n_1 \times n_2 \times ...\; \times n_d}$ is written as:
\begin{equation}
\mathcal{X}_{i_1,...i_d}= {G}_1(i_1) \cdot {G}_2(i_2) \cdots {G}_d(i_d),
\end{equation}
\noindent where ${G}_m(i_m)\in \mathbb{R}^{R_{m-1} \times R_m}$ is the $i_m$th lateral slice of the core $\mathcal{G}_m\in \mathbb{R}^{R_{m-1}\times n_m \times R_m}$, $R_m$s being the TT-rank with $R_0=R_d=1$.

\textit{Computational Issues:} TT decomposition of $\mathcal{X}$ is obtained through a sequence of SVDs. First, $\mathcal{G}_1$ is obtained from SVD of mode-1 matricization of $\mathcal X$ as
\begin{equation}
X^{(1)}=USV^{\top}
\end{equation}
\noindent where $\mathcal{G}_1=U\in\mathbb{R}^{n_1\times \vartriangle_1}$ and $rank(U)=R_1\leq n_1$. Note that, $SV^{\top}\in \mathbb{R}^{R_1 \times n_2 n_3  ... n_d}$. Let $W\in \mathbb{R}^{R_1 n_2 \times n_3 ... n_d}$ be a reshaped version of $SV^{\top}$.  Then, $\mathcal{G}_2\in\mathbb{R}^{R_1\times n_2\times R_2}$ is obtained by reshaping left-singular vectors of $W$, where $W=USV^{\top}$ and $U\in\mathbb{R}^{R_1 n_2\times R_2}$, $rank(U)=R_2\leq rank(W)$ and $SV^{\top}\in\mathbb{R}^{R_2\times n_3...n_d}$.
By repeating this procedure, all of  the core tensors $\mathcal{G}_i$s are obtained by a sequence of SVD decompositions of specific matricizations of $\mathcal{X}$. The storage complexity of TT-rank-R approximation of a $d$th-order tensor  $\mathcal{X}\in \mathbb{R}^{n_{1}\times n_{2}...\times n_{d}}$ with $n_{1}=n_{2}=\ldots=n_{d}=n$ is $\mathcal{O}(dnR^2)$.

It is important to note that the TT format is known as the Matrix Product State (MPS) representation with the Open Boundary Conditions (OBC) in the quantum physics community \cite{orus2014practical}. Some of the advantages of the TT/MPS model over HT are its simpler practical implementation as no binary tree needs to be determined, the simplicity of the computation, computational efficiency (linear in the tensor order).
%%\iffalse

Although TT format has been used widely in signal processing and machine learning, it suffers from a couple of limitations. First, TT model requires rank-1 constraints on the border factors, i.e. they have to be matrices.  Second and most importantly, the multiplications of the TT cores are not permutation invariant requiring the optimization of the ordering using procedures such as mutual information estimation \cite{barcza2011quantum,ehlers2015entanglement}. These drawbacks have been recently addressed by the tensor ring (TR) decomposition \cite{zhao2016tensor,wang2017efficient}. TR decomposition removes the unit rank constraints for the boundary cores and utilizes a trace operation in the decomposition, which removes the dependency on the core order.
\subsection{Tensor Singular Value Decomposition (t-SVD)}
t-SVD is defined for 3rd order tensors based on t-product \cite{kilmer2013third}. Algebra behind t-SVD is different than regular multilinear algebra and  depends on linear operators defined on third-order tensors. In this approach, third-order tensor $\mathcal{X}\in \mathbb{R}^{n_1\times n_2\times n_3}$ is decomposed as
\begin{equation}
\mathcal{X}= \mathcal{U}*\mathcal{S}*\mathcal{V}^T,
\end{equation}
where $\mathcal{U}\in \mathbb{R}^{n_1 \times n_1 \times n_3}$ and $\mathcal{V}\in \mathbb{R}^{n_2 \times n_2 \times n_3}$ are orthogonal tensors with respect to the $'*'$ operation, $\mathcal{S}^{n_1 \times n_2 \times n_3}$ is a tensor whose rectangular frontal slices are diagonal, and the entries in $\mathcal{S}$ are called the singular values of $\mathcal{X}$. $'*'$ denotes the t-product and defined as the circular convolution between mode-$3$ fibers of same size. One can obtain this decomposition by computing matrix
SVDs in the Fourier domain. t-SVD defines the notion of tubal rank, where the tubal rank of $\mathcal{X}$ is defined to be the number of non-zero singular
tubes of $\mathcal{S}$. Moreover, unlike CPD and Tucker models, truncated t-SVD with a given rank can be shown be the optimal approximation in terms of minimizing the Frobenius norm of the error. rank-R approximation of  $\mathcal{X}\in \mathbb{R}^{n_1\times n_2\times n_3}$ can be represented using $\mathcal{O}(Rn_3(n_1 + n_2+1))$ entries.
%%\fi

\subsection{An Empirical Comparison of Different Tensor Decomposition Methods}
\begin{figure}[h t]
	\centering
	\begin{subfigure}{0.4\textwidth}
		\centering
		\includegraphics[width=1\linewidth]{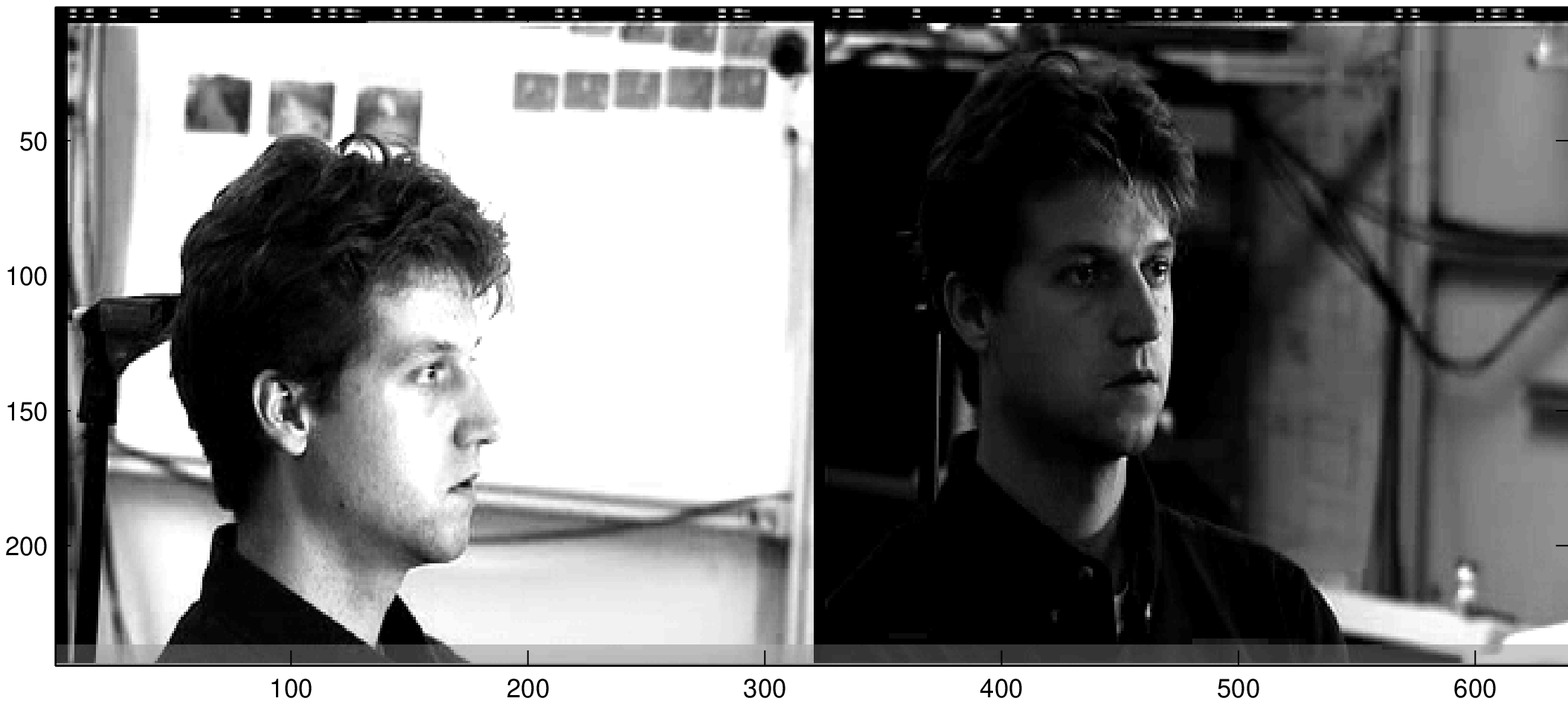}
		\caption{}
		\label{fig:sample-pie}
	\end{subfigure} \hspace{3mm}
	
	\begin{subfigure}{0.27\textwidth}
		\centering
		\includegraphics[width=1\linewidth]{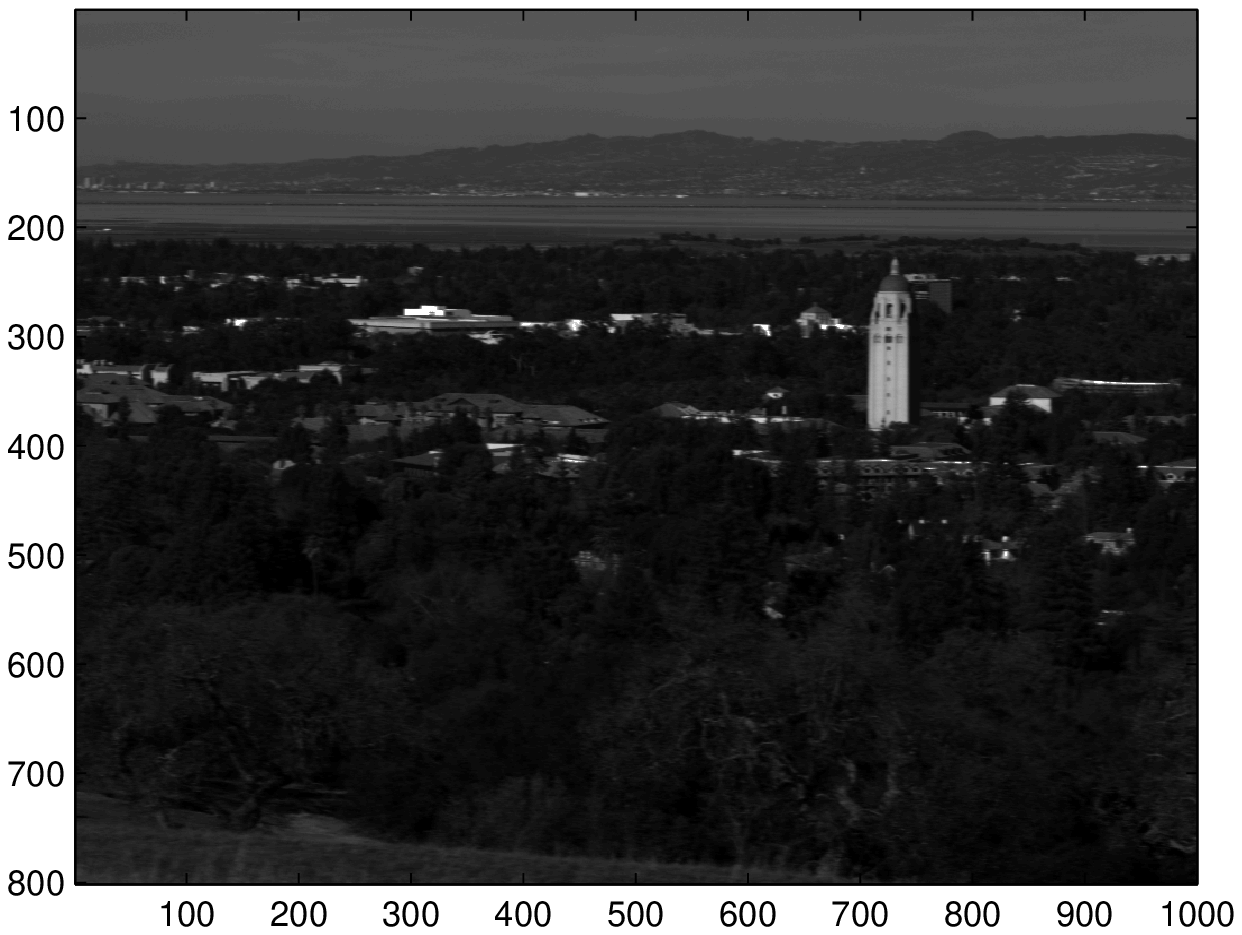}
		\caption{}
		\label{fig:sample-hsi}
	\end{subfigure} \hspace{3mm}
	\begin{subfigure}{0.3\textwidth}
		\includegraphics[width=1\linewidth]{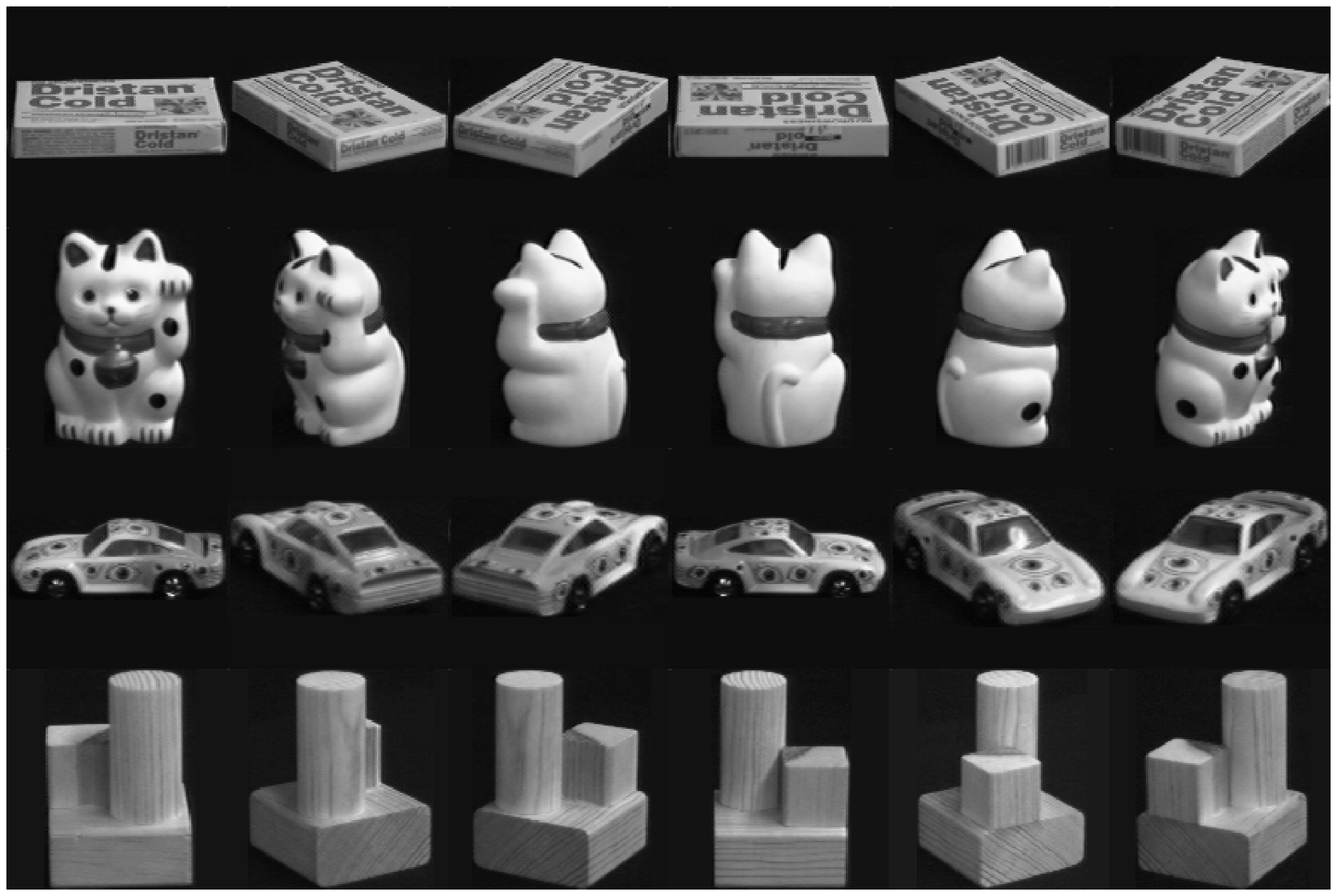}
		\caption{}
		\label{fig:sample-coil}
	\end{subfigure}
	
	\caption{Sample images from data sets used in experiments. (a) The PIE data set. (b) The Hyperspectral Image. (c) The COIL-100 data set.}
	\label{fig:sample}
\end{figure}

In this section the CANDECOMP/PARAFAC, Tucker (HOOI), HoSVD, HT, and TT decompositions are compared in terms of data reduction rate and normalized reconstruction error. The data sets used for this purpose are:\\
(1) {\it PIE data set:} This database contains $138$ images taken from one individual under different illumination conditions and from $6$ different angles \cite{PIE}. All $244 \times 320$ images of the individual form a $3$-mode tensor $\mathcal{X} \in \mathbb{R}^{244 \times 320 \times 138}$.\\
(2) {\it Hyperspectral Image (HSI) data set:} This database contains $100$ images taken at $148$ wavelengths \cite{HSI}. Images consist of $801 \times 1000$ pixels, forming a $3$-mode tensor $\mathcal{X} \in \mathbb{R}^{801 \times 1000 \times 148}$.\\
(3) {\it COIL-100 data set:} This database includes $7200$ images taken from $100$ objects \cite{COIL-100}. Each object was imaged at $72$ different angles separated by $5$ degrees, resulting in $72$ images per object, each one consisting of $128 \times 128$ pixels. The original database is a $128 \times 128 \times 7200$ tensor which was reshaped as a $4$-mode tensor $\mathcal{X} \in \mathbb{R}^{128 \times 128 \times 72 \times 100}$ for the experiments.\\
Sample images from the above data sets can be viewed in Figure \ref{fig:sample}.

Several software packages were used to generate the results.  The Tucker (HOOI) and CPD methods were evaluated using the {\it TensorLab} package \cite{Tensorlab}.  For CPD, the structure detection and exploitation option was disabled in order to avoid complications on the larger datasets ({\it COIL-100} and {\it HSI}).\footnote{To disable the structure exploitation option, the field {\it options.ExploitStructure} was set to {\it false} in cpd.m. The large-scale memory threshold was also changed from $2$ GB to $16$ GB in mtkrprod.m to prevent the large intermediate solutions from causing memory issues.} For CPD, the number of rank-$1$ tensors in the decomposition, $R$, is given as the input parameter to evaluate the approximation error with respect to compression rate.  For HOOI, the input is a cell array of factor matrices for different modes that are used as initialization for the main algorithm. To generate the initial factor matrices, HoSVD  was computed with the {\it TP Tool} \cite{TPtool}. The original version of the code was slightly modified in a way that the input parameter to HoSVD is the threshold $0 \leq \tau \leq 1$ defined in (\ref{Equ:threshold}), where $n_0$ is the minimum number of singular values that allow the inequality to hold for mode $i$. The same threshold value, $\tau$, was chosen for all modes.
\begin{equation}
\frac{\sum\limits_{k=1}^{n_0} \sigma_k}{\sum\limits_{k=1}^{n_i} \sigma_k} \geq \tau.
\label{Equ:threshold}
\end{equation}

The {\it Tensor-Train Toolbox} \cite{TT} was used to generate results for the Tensor-Train (TT) method.  The input parameter for TT is the accuracy (error) with which the data should be stored in the Tensor-Train format.  Similarly, the {\it Hierarchical Tucker Toolbox} \cite{HT} was used to generate results for the Hierarchical Tucker (HT) method.  The input parameter for HT is the maximal hierarchical rank.
%The different ranges of input parameters for each method have been summarized in Table (\ref{table:range}),
The different ranges of input parameters for each method have been summarized in Table \ref{table:range}. These parameters have been chosen to yield comparable reconstruction errors and compression rates across different methods.

\begin{figure}[h t]
	\centering
	
	\begin{subfigure}{0.35\textwidth}
		\centering
		\includegraphics[width=1\linewidth]{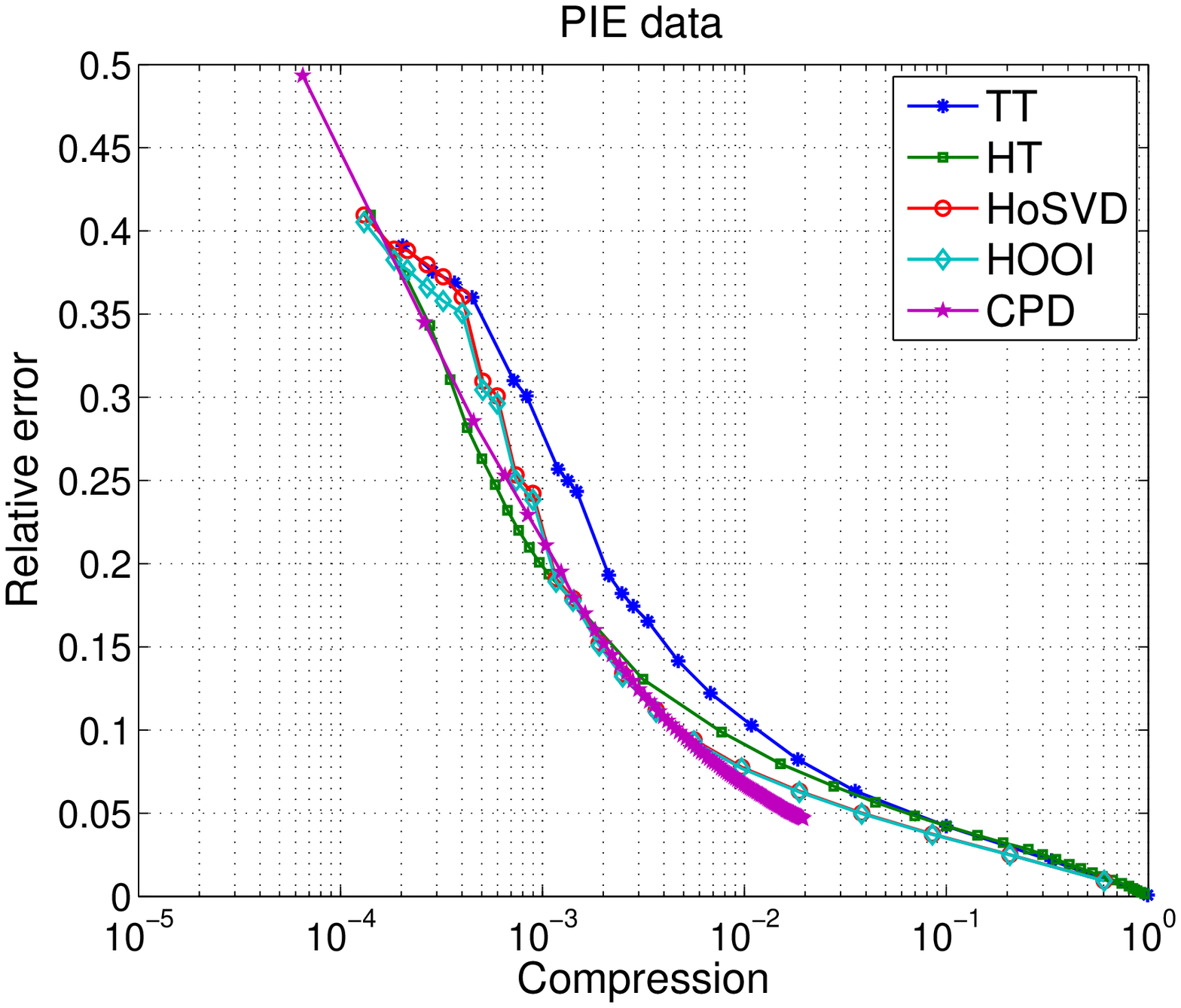}
		\caption{}
		\label{fig:result-pie}
	\end{subfigure}
	
	\begin{subfigure}{0.35\textwidth}
		\centering
		\includegraphics[width=1\linewidth]{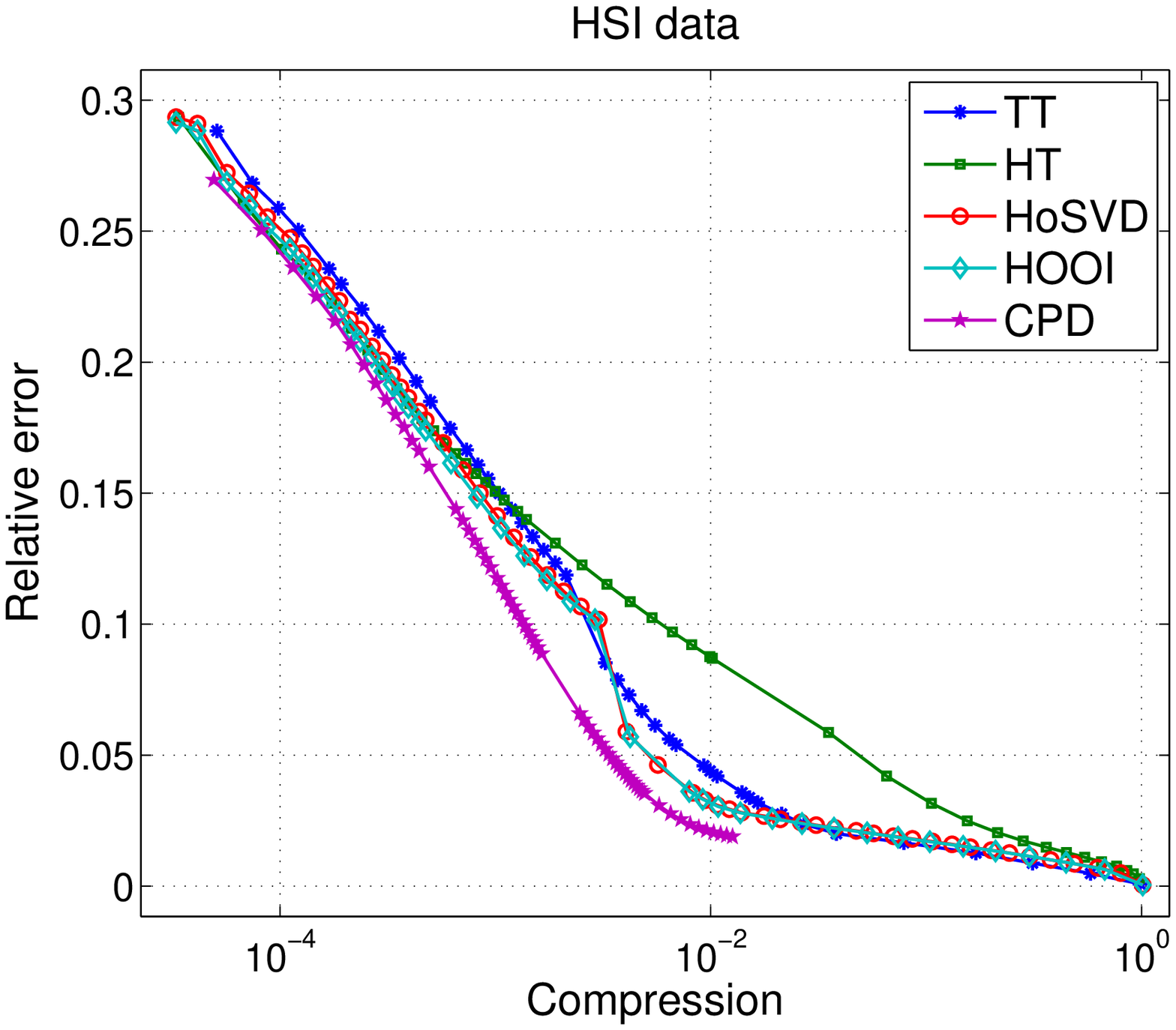}
		\caption{}
		\label{fig:result-hsi-all}
	\end{subfigure}\hspace{-4mm}
	\begin{subfigure}{0.35\textwidth}
		\includegraphics[width=1\linewidth]{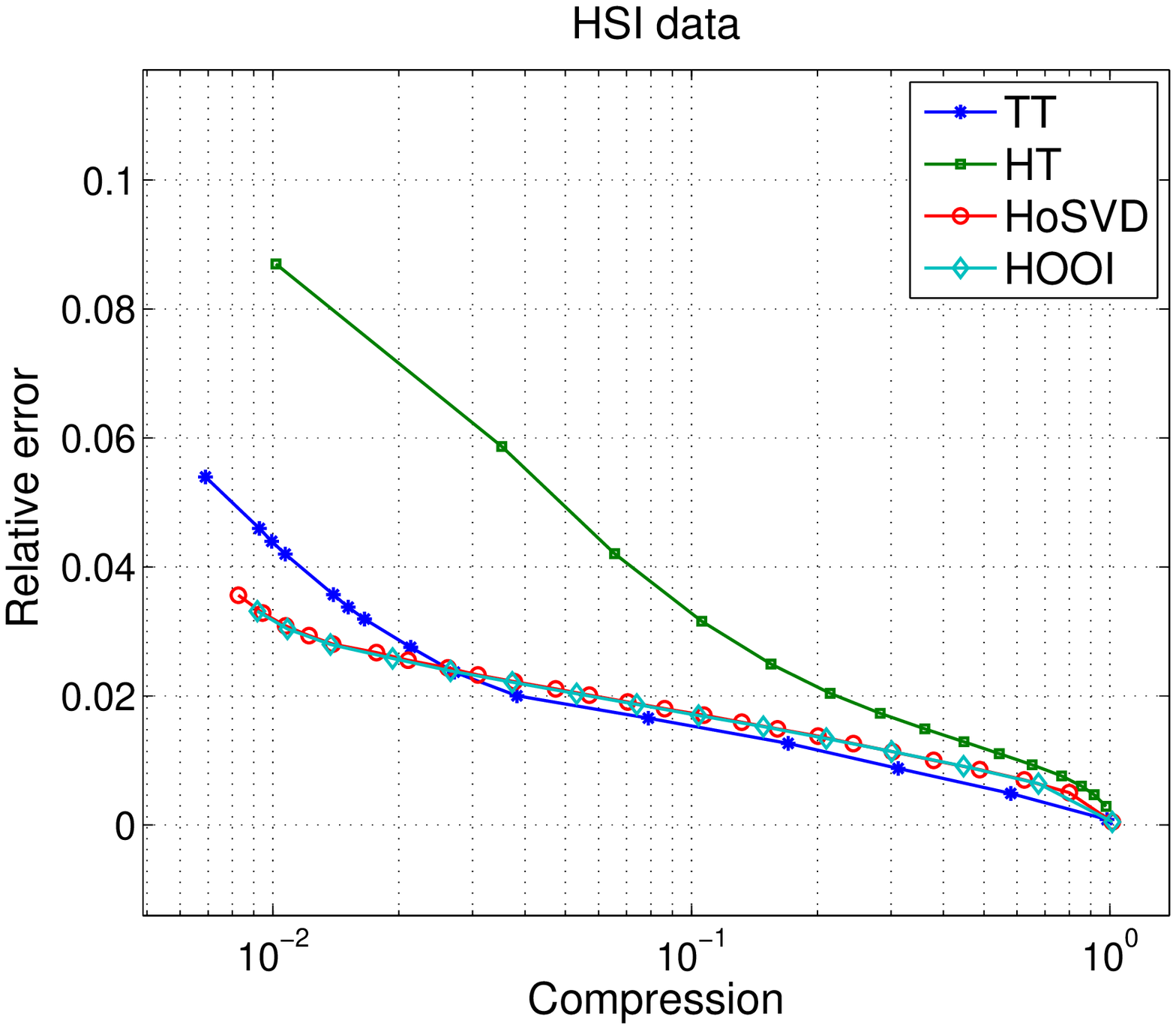}
		\caption{}
		\label{fig:result-hsi-nocpd}
	\end{subfigure}
	
	\begin{subfigure}{0.35\textwidth}
		\centering
		\includegraphics[width=1\linewidth]{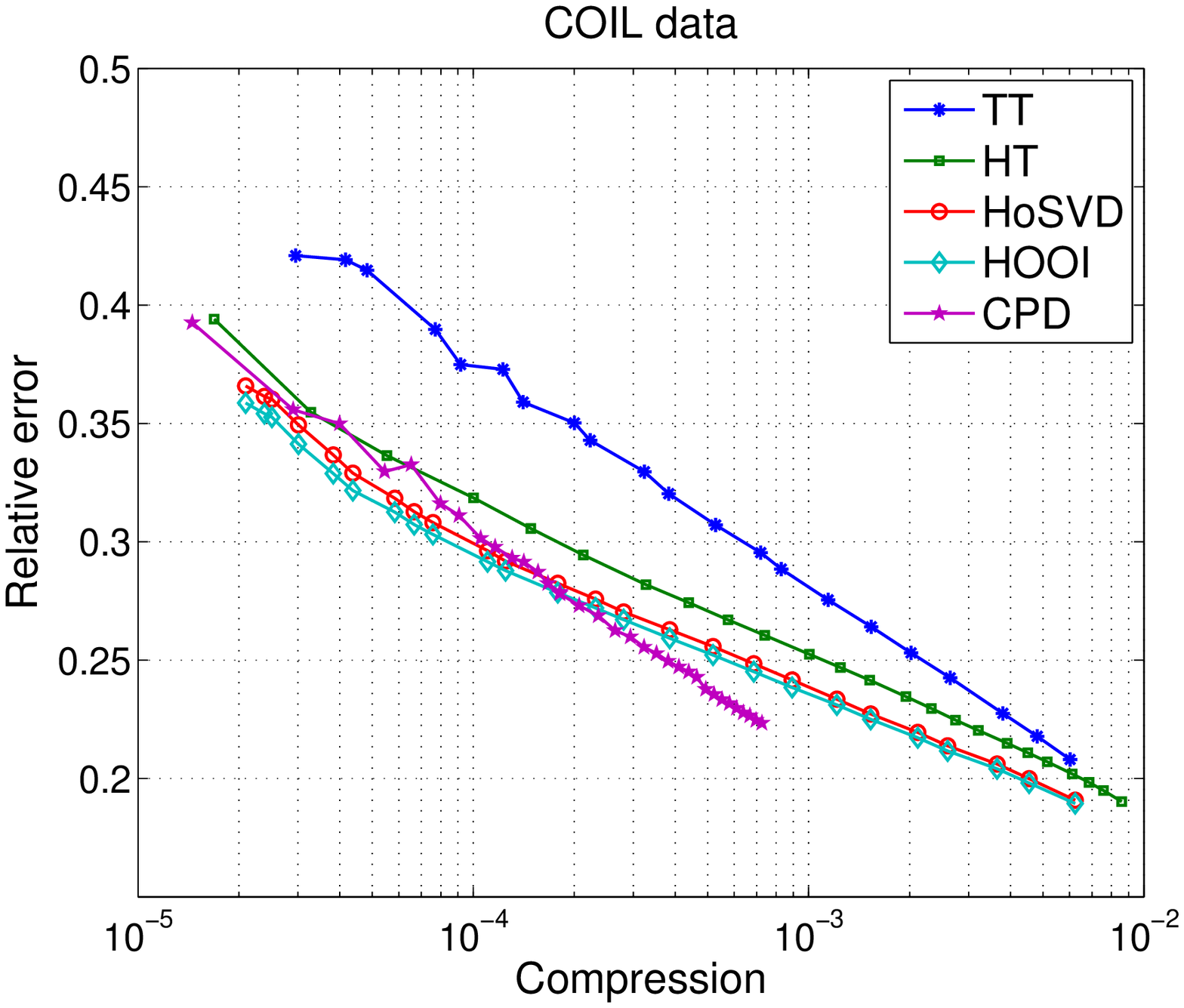}
		\caption{}
		\label{fig:result-coil-all}
	\end{subfigure}\hspace{-4mm}
	\begin{subfigure}{0.35\textwidth}
		\includegraphics[width=1\linewidth]{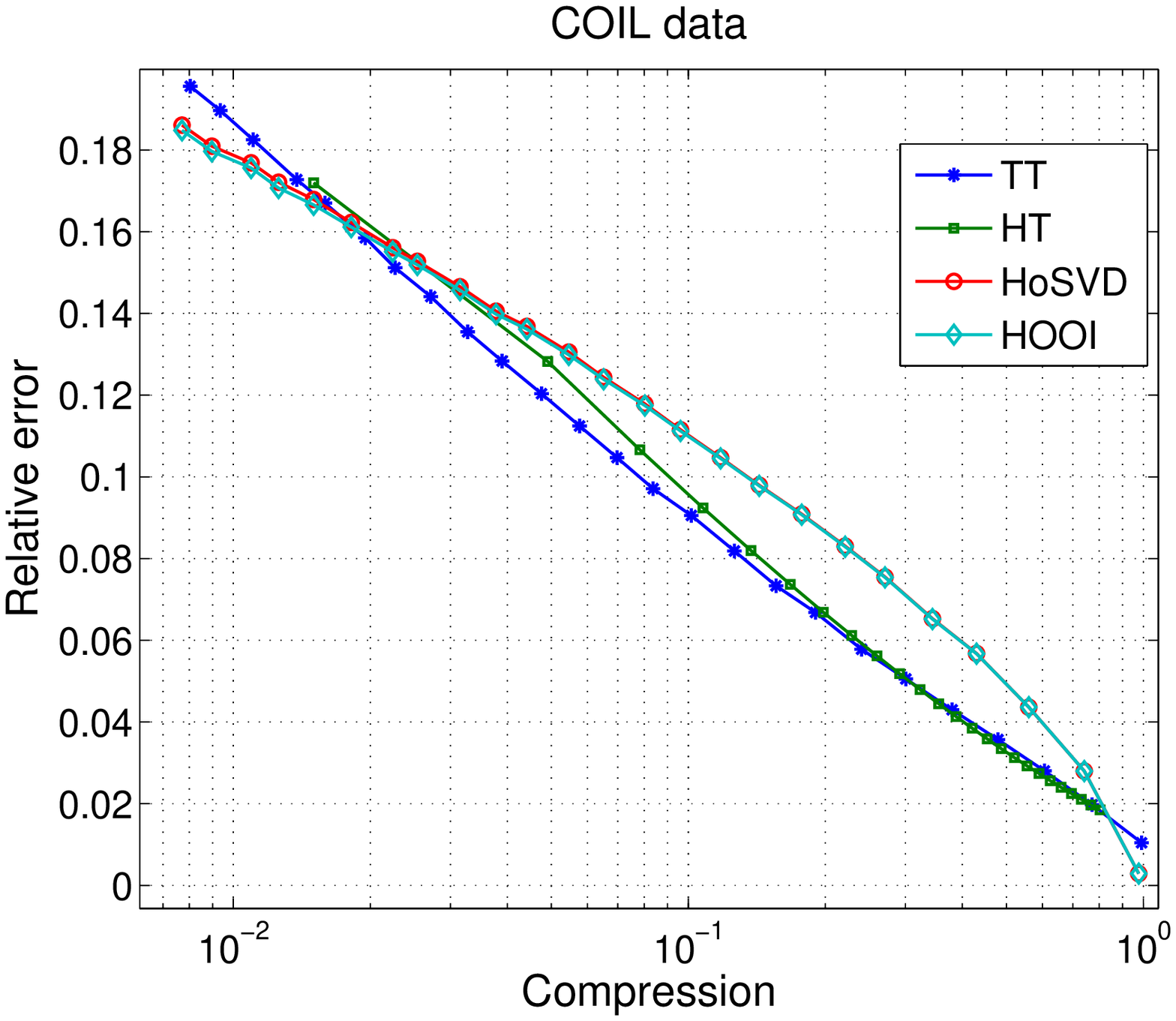}
		\caption{}
		\label{fig:result-coil-nocpd}
	\end{subfigure}
	
	\caption{Experimental results. Compression versus error performance of CPD, TT, HT, HOOI and the HoSVD on:  (a) the PIE data set. (b) HSI data results for all $5$ methods. (c) HSI data results without CPD for higher compression values. (d) COIL-100 data results for all $5$ methods. (e) COIL-100 data without CPD for higher compression values.}
	\label{fig:results}
\end{figure}

The experimental results are given in Figure \ref{fig:results}. Here, compression is defined as the ratio of size of the output data to the size of input data.\footnote{No other data compression methods were used to help reduce the size of the data. Data size was compared in terms of bytes needed to represent each method's output versus its input. When the same precision is used for all data, this is equal to measuring the total number of elements of the input and output arrays.} The relative error was calculated using %\eqref{Equ:error}.
\begin{equation}
E=\frac{\|input-output\|}{\|input\|},
\label{Equ:error}
\end{equation}
where {\it input} denotes the input tensor and {\it output} denotes the (compressed) tensor reconstructed using the given method.

% Good to have this for our records, but this is possibly too much detail for the paper.  Let's keep it commented.
\begin{table}[h t]
	\centering
	\caption{Range of input parameter values used for the different methods to generate the results shown in Figure \ref{fig:results}.}
	\label{table:range}
	\begin{tabular}{|c|c|c|c|c|c|} \hline
		Data                        & PIE       & \multicolumn{2}{c|}{HSI}     & \multicolumn{2}{c|}{COIL-100} \\ \hline
		Figure                      & 5(a)      & 5(b)        & 5(c)          & 5(d)         & 5(e)          \\ \hline
		TT (error)                    & 0.001-0.5 & 0.001-0.5   & 0.001-0.0555  & 0.21-0.5     & 0.0831-0.2    \\ \hline
		HT (max. H. rank)              & 2-300     & 2-960       & 100-960       & 1-80         & 100-3500      \\ \hline
		HoSVD (threshold $\tau$)    & 0.2-0.99  & 0.15-0.9999 & 0.8841-0.9999 & 0.4-0.76     & 0.77-0.999    \\ \hline
		HOOI  (threshold $\tau$)   & 0.2-0.99  & 0.15-0.9999 & 0.8864-0.9999 & 0.4-0.76     & 0.77-0.999    \\ \hline
		CPD (rank $R$) & 1-300     & 3-767       & -             & 4-200        & -  \\  \hline
	\end{tabular}
\end{table}

The following observations can be made from Figure~\ref{fig:results}.  CPD has difficulty converging when the input rank parameter, $R$, is large especially for larger data sets.  However, it provides the best compression performance when it converges. For the PIE data set, HT and specifically TT do not perform well for most compression ranges. However, HT outperforms the other methods with respect to approximation error for compression rates below $10^{-2}$. For the HSI data set, TT and HT again do not perform very well, particularly for compression values around $10^{-2}$. HT continues to perform poorly for larger compression rates as well. For the COIL-$100$ data set, TT and HT generate very good results for compression rates above $10^{-2}$, but fail to do so for lower compression rates. This is especially the case for TT.  It should be noted that the largest number of modes for the tensors considered in this paper is only $4$. Tensor network models such as TT and HT are expected to perform better for higher number of modes.  HoSVD and HOOI provide very similar results in all data sets, with HOOI performing slightly better. They generally appear to provide the best compression versus error results of all the methods in regimes where CPD does not work well, with the exception of the COIL dataset where HT and TT outperform them at higher compression rates.

%%%%%%%%%%%%%%%%%%%%%%%%%%%%%%%%%%%%%%%%%%%
% TENSOR PCA
%%%%%%%%%%%%%%%%%%%%%%%%%%%%%%%%%%%%%%%%%%%

\section{Tensor PCA Approaches}
\label{sec:TensorPCA}
In this section, we will review some of the recent developments in supervised learning using tensor subspace estimation methods. The methods that will be discussed in this section aim to employ tensor decomposition methods for tensor subspace learning to extract low-dimensional features. We will first start out by giving an overview of Trivial PCA based on tensor vectorization. Next, we will review PCA-like feature extraction methods extended for Tucker, CPD, TT and HT tensor models. Finally, we will give an overview of the tensor embedding methods for machine learning applications.
\subsection{Trivial PCA for Tensors Based on Implicit Vectorization}
\label{sec:TrivialPCA}

Some of the first engineering methodologies involving PCA for $d$th-order ($d\geq 3$) tensor data were developed in the late $1980$'s in order to aid in facial recognition, computer vision, and image processing tasks (see, e.g., \cite{Sirovich_Low-dimensional_1987,Turk_Eigenfaces_1991,Yang_Face_2000,Merhi_Face_2012} for several variants of such methods).  In these applications preprocessed pictures of $m$ individuals were treated as individual $2$-mode tensors.  In order to help provide additional information each individual might further be imaged under several different conditions (e.g., from a few different angles, etc.).  The collection of each individual's images across each additional condition's mode (e.g., camera angle) would then result in a $d$th-order ($d\geq 3$) tensor of image data for each individual.  The objective would then be to perform PCA across the individuals' face image data (treating each individual's image data as a separate data point) in order to come up with a reduced face model that could later be used for various computer vision tasks (e.g., face recognition/classification).

Mathematically these early methods perform implicitly vectorized PCA on $d$-mode tensors $\mathcal{A}_1, \dots, \mathcal{A}_m \in \mathbb{R}^{n_1 \times n_2 \times \dots \times n_d}$ each of which represents an individual's image(s).  Assuming that the image data has been centered so that $\sum^{m}_{j=1} \mathcal{A}_j = 0$, this problem reduces to finding a set of $R < m$ orthonormal ``eigenface'' basis tensors $\mathcal{B}_1, \dots, \mathcal{B}_R \in \mathbb{R}^{n_1 \times n_2 \times \dots \times n_d}$ whose {\it span} $S$,
$$S := \left \{ \sum^R_{j = 1}\alpha_j \mathcal{B}_j ~\big|~ \bm{\alpha} \in \mathbb{R}^R \right\} \subset \mathbb{R}^{n_1 \times n_2 \times \dots \times n_d},$$
minimizes the error
\begin{equation}
E_{\rm PCA}(S) := \sum^m_{j = 1} \min_{\mathcal{X}_j \in S} \| \mathcal{A}_j - \mathcal{X}_j \|^2,
\end{equation}
where $\mathcal{X}_{j}$ provides approximation for each of the original individual's image data, $\mathcal{A}_j$, in compressed form via a sum
\begin{equation}
\mathcal{A}_j \approx \mathcal{X}_j = \sum^R_{\ell = 1} \alpha_{j,\ell} \mathcal{B}_\ell
\end{equation}
for some optimal $\alpha_{j,1}, \dots, \alpha_{j,R} \in \mathbb{R}$.

It is not too difficult to see that this problem can be solved in vectorized form by (partially) computing the Singular Value Decomposition (SVD) of an $\mathbb{R}^{n_1 n_2 \dots n_d \times m}$ matrix whose columns are the vectorized image tensors ${\bf a}_1, \dots, {\bf a}_m \in \mathbb{R}^{n_1 n_2 \dots n_d}$.  As a result one will obtain a vectorized ``eigenface'' basis ${\bf b}_1, \dots, {\bf b}_r \in \mathbb{R}^{n_1 n_2 \dots n_d}$ each of which can then be reshaped back into an image tensor $\in \mathbb{R}^{n_1 \times n_2 \times \dots \times n_d}$.  Though conceptually simple this approach still encounters significant computational challenges.  In particular, the total dimensionality of each tensor, $n_1 n_2 \dots n_d$, can be extremely large making both the computation of the SVD above very expensive, and the storage of the basis tensors $\mathcal{B}_1, \dots, \mathcal{B}_R$ inefficient.  The challenges involving computation of the SVD in such situations can be addressed using tools from numerical linear algebra (see, e.g., \cite{Halko_Finding_2011,Iwen_Distributed_2016}).  The challenges involving efficient storage and computation with the high dimensional tensors $\mathcal{B}_j$ obtained by this (or any other approach discussed below) can be addressed using the tensor factorization and compression methods discussed in Section~\ref{sec:TensorDecomps}.

\subsection{Multilinear Principal Component Analysis (MPCA)}

The first Tensor PCA approach we will discuss, MPCA \cite{Lu_MPCA_2008,Xu_Reconstruction_2008}, is closely related to the Tucker decomposition of a $d$th-order tensor \cite{Tucker_Extension_1964,Lathauwer_Multilinear_2000}.  MPCA has been independently discovered in several different settings over the last two decades. The first MPCA variants to appear in the signal processing community focused on $2$nd-order tensors, e.g. 2DPCA, with the aim of improving image classification and database querying applications \cite{Yang_Two-dimensional_2004,Ye_GPCA_2004,Ye_Generalized_2005,Liu_Generalized_2010}.  The methods were then later generalized to handle tensors of any order \cite{Lu_MPCA_2008,Xu_Reconstruction_2008}.  Subsequent work then tailored these general methods to several different applications including variants based on non-negative factorizations for audio engineering \cite{Panagakis_Non-Negative_2010}, weighted versions for EEG signal classification \cite{Washizawa_Tensor_2010}, online versions for tracking \cite{Wang_Tracking_2011}, variants for binary tensors \cite{Mazgut_Dimensionality_2014}, and incremental versions for streamed tensor data \cite{Sun_Incremental_2008}.

MPCA performs feature extraction by determining a multilinear projection that captures
most of the original tensorial input variation, similar to the goals of PCA for vector type data. The solution is iterative in nature and it proceeds by
decomposing the original problem to a series of multiple projection subproblems. Mathematically, all of the general MPCA methods \cite{Yang_Two-dimensional_2004,Ye_GPCA_2004,Ye_Generalized_2005,Liu_Generalized_2010,Lu_MPCA_2008,Xu_Reconstruction_2008} aim to solve the following problem (see, e.g., \cite{Lu_Multilinear_2013} for additional details):  Given a set of higher-order centered data as in \ref{sec:TrivialPCA}, $\mathcal{A}_1, \dots, \mathcal{A}_m \in \mathbb{R}^{n_1 \times n_2 \times \dots \times n_d}$,  MPCA aims to find $d$ low-rank orthogonal projection matrices $U^{(j)} \in \mathbb{R}^{n_j \times R_j}$ of rank $R_j \leq n_j$ for all $j \in [d]$ such that the projections of the tensor objects to the lower dimensional subspace, $S := \left \{ \mathcal{B} \times_1 U^{(1)} \dots \times_d U^{(d)}~\big|~ \mathcal{B} \in \mathbb{R}^{n_1 \times n_2 \times \dots \times n_d} \right\} \subset \mathbb{R}^{R_1 \times R_2 \times \dots \times R_d}$, minimize the error
\begin{equation}
E_{\rm MPCA}(S) := \sum^m_{j = 1} \min_{\mathcal{X}_j \in S} \| \mathcal{A}_j - \mathcal{X}_j \|^2 = \min_{\mathcal{U}^{(1)},\mathcal{U}^{(2)},\ldots, \mathcal{U}^{(d)}}\sum^m_{j = 1} \| \mathcal{A}_j - \mathcal{A}_j \times_1 U^{(1),T} \dots \times_d U^{(d),T} \|^2.
\label{equ:MPCAerror}
\end{equation}
 This error can be equivalently written in terms of the tensor object projections defined as $\mathcal{X}_{j}=\mathcal{A}_{j}\times_{1}U^{(1),T}\times_{2}\ldots\times_{d}U^{(d),T}$, where $\mathcal{X}_{j}\in \mathbb{R}^{R_{1}\times R_{2} \times \ldots \times R_{d}}$ capture most of the variation observed in the original
tensor objects. When the variation is quantified as the total scatter of the projected tensor objects, then \cite{Lu_MPCA_2008,Lu_Multilinear_2013} have shown that for given all the other
projection matrices $U^{(1)},\ldots,U^{(n-1)},U^{(n+1)},\ldots,U^{(d)}$, the matrix $U˜^{(n)}$ consists of the $R_{n}$ eigenvectors
corresponding to the largest $R_{n}$ eigenvalues of the matrix $\Phi_{n}=\sum_{k=1}^{m}A_{k,(n)}U_{\Phi_{n}}U_{\Phi_{n}}^{T}A_{k,(n)}^{T}$ where $U_{\Phi_{n}}=(U^{(n+1)}\otimes U^{(n+2)}\otimes \ldots \otimes U^{(d)}\otimes U^{(1)}\otimes U^{(2)} \ldots \otimes U^{(n-1)})$. Since the optimization of
$U^{(n)}$ depends on the projections in other modes and there is no closed form solution to this maximization
problem.

A subspace minimizing \eqref{equ:MPCAerror} can be approximated using Alternating Partial Projections (APP).  This iterative approach simply fixes $d-1$ of the current mode projection matrices before optimizing the single remaining free mode's projection matrix in order to minimize \eqref{equ:MPCAerror} as much as possible.  Optimizing over a single free mode can be accomplished exactly by computing (a partial) SVD of a matricized version of the tensor data.  In the next iteration, the previously free mode is then fixed while a different mode's projection matrix is optimized instead, etc.. Although there are no theoretical guarantees for the convergence of MPCA, as the total scatter is a
non-decreasing function MPCA converges fast as shown through multiple empirical studies. It has also been shown that MPCA reduces to PCA for $d=1$ and to 2DPCA for $d=2$.

Recently, MPCA has been generalized under a unified framework called generalized tensor PCA (GTPCA) \cite{inoue2016generalized}, which includes MPCA, robust MPCA \cite{inoue2009robust}, simultaneous low-rank approximation of tensors (SLRAT) and robust SLRAT \cite{inoue2009robustslrat} as its special cases. This generalization is obtained by considering different cost functions in tensor approximation, i.e. change in \eqref{equ:MPCAerror}, and different ways of centering the data.

There are a couple of important points that distinguish MPCA from HoSVD/Tucker model discussed in Section~\ref{sec:TensorDecomps}. First, the goal of MPCA is to find a common low-dimensional subspace across multiple tensor objects such that the resulting projections capture the maximum total variation of the input tensorial space. HoSVD, on the other hand, focuses on the low-dimensional representation of a single tensor object with the goal of obtaining a low-rank approximation with a small reconstruction error,  i.e. compression. Second, the issue of centering has been ignored in tensor decomposition as the focus is predominantly on tensor approximation and reconstruction. For the approximation/reconstruction problem, centering is
not essential, as the (sample) mean is the main focus of attention. However, in machine learning applications
where the solutions involve eigenproblems, non-centering
can potentially affect the per-mode eigendecomposition and lead to a solution that captures the variation
with respect to the origin rather than capturing the true variation of the data. Finally, while the initialization of $U^{(i)}$s are usually done using HoSVD, where the first $R_{n}$ columns of the full Tucker decomposition are kept, equivalent to the HoSVD-based
low-rank tensor approximation, the final projection matrices are different than the solution of HoSVD.
\subsection{Tensor Rank-One Decomposition (TROD)}

As mentioned above, MPCA approximates the given data with subspaces that are reminiscent of the Tucker decomposition (i.e., subspaces defined in terms of $j$-mode products).  Similarly, the TROD approach \cite{Shashua_Linear_2001} formulates PCA for tensors in terms of subspaces that are reminiscent of the PARAFAC/CANDECOMP decomposition \cite{Harshman_PARAFAC_1970,Kruskal_Three_1977,Bro_PARAFAC_1997,Faber_Recent_2003} (i.e., subspaces whose bases are rank-1 tensors).  Given higher-order data $\mathcal{A}_1, \dots, \mathcal{A}_m \in \mathbb{R}^{n_1 \times n_2 \times \dots \times n_d}$ TROD aims to find $dR$ vectors ${\bf u}_{j}^{(1)}, \dots, {\bf u}_{j}^{(R)} \in \mathbb{R}^{n_j}$ for all $j \in [d]$  such that the resulting subspace
$$S := \left \{  \sum^R_{k = 1} \alpha_k \bigotimes^d_{j=1} {\bf u}_{j}^{(k)} ~\Big|~ \bm{\alpha} \in \mathbb{R}^R \right\} \subset \mathbb{R}^{n_1 \times n_2 \times \dots \times n_d}$$
minimizes the error
\begin{equation}
E_{\rm TROD}(S) := \sum^m_{j = 1} \min_{\mathcal{X}_j \in S} \| \mathcal{A}_j - \mathcal{X}_j \|^2.
\label{equ:TRODerror}
\end{equation}
A subspace minimizing \eqref{equ:TRODerror} can again be found by using a greedy least squares minimization procedure to iteratively compute each ${\bf u}_{j}^{(k)}$ vector while leaving all the others fixed. If the underlying tensor objects do not conform to a low CPD rank model, this method suffers from high computational complexity and slow convergence rate \cite{Shashua_Linear_2001}.
\subsection{Tensor Train PCA (TT-PCA)}

As Tensor Train (TT) and hierarchical Tucker representation have been shown to provide alleviations to the storage requirements of Tucker representation, recently, these decomposition methods have been extended for feature extraction and machine learning applications \cite{bengua2017matrix,wang2018principal,chaghazardi2017sample}. Given a set of tensor data $\mathcal{A}_{1},\ldots,\mathcal{A}_{m}\in \mathbb{R}^{n_{1}\times n_{2}\times \ldots \times n_{d}}$,  the objective is to find $3$-mode tensors $\mathcal{U}_{1},\mathcal{U}_{2},
\ldots,\mathcal{U}_{n}$ such that the distance of points from the subspace is minimized, i.e.
\begin{equation}
E_{\rm TT}:= \min_{\mathcal{U}_{i},i\in[n],A}\|\mathbf{L}(\mathcal{U}_{1}\ldots\mathcal{U}_{n})A-D\|^{2},
 \end{equation}
 where $\mathbf{L}$ is the left unfolding operator resulting in a matrix that takes the first two mode indices as row indices and the third mode indices as column indices, $D$ is a matrix that concatenates  the vectorizations of the sample points and $A$ is the representation matrix. Wang et al. \cite{wang2018principal} propose an approach based on successive SVD-algorithm for computing TT decomposition followed by thresholding the singular values. The obtained subspaces across each mode are left-orthogonal which provides a simple way to find the representation of a data point in the subspace $\mathbf{L}(\mathcal{U}_{1}\ldots\mathcal{U}_{n})$ through a projection operation.

A similar TT-PCA algorithm was introduced in \cite{bengua2017matrix} with the main difference being the ordering of the training samples to obtain the $d+1$ mode tensor, $\mathcal{X}$. Unlike \cite{wang2018principal} which places the samples in the last mode of the tensor, in this approach first all modes of the tensor are permuted such that $n_{1}\geq n_{2} \ldots \geq n_{i-1}$ and $n_{i}\leq \ldots \leq n_{d}$. The elements of $\mathcal{X}$ can then be presented in a mixed-canonical form of the tensor train decomposition, i.e. through a product of left and right common factors which satisfy left- and right-orthogonality conditions. By optimizing the positions of tensor modes, one can obtain a reduced rank representation of the tensor and extract features. The implementation of the algorithm relies on successive sequences of SVDs just as in \cite{wang2018principal} followed by thresholding. Finally, subspaces across each mode are extracted similar to \cite{wang2018principal}.
\subsection{Hierarchical tensor PCA (HT-PCA)}
Based on the Hierarchical Tucker (HT) decomposition, one can construct tensor subspaces in a hierarchical manner. Given a set of tensor data $\mathcal{A}_{i}\in \mathbb{R}^{n_{1}\times n_{2}\times \ldots \times n_{d}}, i\in [m]$, a dimension tree $T$ and the ranks corresponding to the nodes in the tree, the problem is to find the best HT subspace, i.e. $U_{t}\in \mathbb{R}^{n_{t}\times R_{t}}$ and the transfer matrices, $B_{t}$ that minimizes the mean square error. As estimating both the dimension tree and the subspaces is a NP hard problem, the current applications have considered balanced tree and TT-tree using a suboptimal algorithm \cite{chaghazardi2017sample}. The proposed algorithm is a variation of the Hierarchical SVD computing HT representation as a single tensor object. The algorithm takes the collection of $m$ tensors and computes the hierarchical space using the given tree. The subspaces corresponding to each node are computed using a truncated SVD on the node unfolding and the transfer tensors are computed using the projections on the tensor product of subspace of the node's children. Empirical results indicate that TT-PCA performs better than HT-PCA in terms of classification error for a given number training samples \cite{chaghazardi2017sample}.
\subsection{Tensor Embedding Methods}
Over the past decade, embedding methods developed for feature extraction and dimensionality reduction in various machine learning tasks have been extended to tensor objects \cite{he2006tensor,dai2006tensor,li2008discriminant,liu2010tensor,wang2017tensor}. These methods take data directly in the form of tensors and allow the relationships between dimensions of tensor representation to be efficiently characterized. Moreover, these methods estimate the intrinsic local geometric and topological properties of the manifold embedded in a tensor space. Some examples include tensor neighborhood preserving embedding (TNPE),  tensor locality preserving projection (TLPP) and tensor local discriminant embedding (TLDE) \cite{dai2006tensor,li2008discriminant}. In these methods, optimal projections that preserve the local topological structure of the manifold embedded in a tensor space are determined through iterative algorithms. The early work in \cite{dai2006tensor} has focused on the Tucker model to define the optimal projections, while more recently TNPE has been extended to the TT model (TTNPE) \cite{wang2017tensor}.

\subsection{An Empirical Comparison of Two Tensor PCA Approaches}
In this section, we evaluate the performance of Tensor PCA based on two tensor models; Tucker and Tensor Train resulting in MPCA \cite{Lu_Multilinear_2013} and TT-PCA \cite{bengua2017matrix} approaches as these are currently the two methods that are widely used for supervised learning. To assess the performance of these methods, binary classification is performed on three data sets, including the {\it COIL-100}, {\it EYFDB}\footnote{Extended Yale Face Database B} \cite{GeBeKr01} and {\it MNIST}\footnote{Modified National Institute of Standards and Technology} \cite{MNIST} databases.

{\it COIL-100:} This data set was reshaped to its original structure, i.e., a $3$-dimensional tensor $\mathcal{X} \in \mathbb{R}^{128 \times 128 \times 7200}$ with $72$ samples per class available. Objects with labels $65$ and $99$ were picked for the experiment.

{\it EYFDB:} This data set is a $4$-dimensional tensor including images from $28$ subjects with different pose angles and under various illumination conditions. Each subject is imaged at $9$ pose angles, and at each angle, $64$ lighting conditions are used to generate images of size $128 \times 128$. This leads to a tensor $\mathcal{X} \in \mathbb{R}^{128 \times 128 \times 64 \times 252}$ containing samples from all $28$ classes. However, with this data structure, only $9$ samples are available in each class, which makes the performance analysis of classification difficult. Therefore, the data were reshaped to $\mathcal{X} \in \mathbb{R}^{128 \times 128 \times 9 \times 1792}$ providing $64$ samples per class. Subjects with labels $3$ and $25$ were selected for the experiment.

{\it MNIST:} This data set contains $20 \times 20$ pixel images of handwritten digits from $0$ to $9$, with each digit (class) consisting of $7000$ samples, resulting in a $3$-mode tensor $\mathcal{X} \in \mathbb{R}^{20 \times 20 \times 70000}$. Digits $5$ and $8$ were chosen for classification.

In each case, two classes were picked randomly, and the nearest neighbor classifier was used for evaluating the classifier accuracy. For $n_t$ training samples and $n_h$ holdout (test) samples, the training ratio $\gamma_t=n_t/\left(n_t+n_h\right)$ was varied, and for each value of $\gamma_t$, the threshold $\tau$ in (\ref{Equ:threshold}) was also varied to see the effect of number of training samples and compression ratio on the Classification Success Rate (CSR). In addition, for each data set, classification was repeated multiple times by randomly selecting the training samples. The classification experiments  for {\it COIL-100}, {\it EYFDB} and {\it MNIST} data were repeated $10$, $50$ and $10$ times, respectively. The average and standard deviation of CSR values can be viewed in Tables \ref{table:class-coil}-\ref{table:class-MNIST}. As can be observed, for the {\it COIL-100} and {\it EYFDB} data, MPS performs better compared to MPCA across different sizes of the training set and compression ratios, especially for smaller training ratios $\gamma_t$. In the case of {\it MNIST} data, the performance of the two approaches are almost identical, with MPCA performing slightly better. This may be due to the fact that MNIST fits the Tucker model better as the variations across the samples within the class can be approximated well with a few number of factor matrices across each mode. These results illustrate that with tensor network models, it is possible to obtain high compression and learn discriminative low-dimensional features, simultaneously.

\begin{table}
	\parbox{.45\linewidth}{
		\centering
		\caption{Binary Classification results (percent) for {\it COIL-100}. Objects with labels $65$ and $99$ were used.}
		\label{table:class-coil}
		\setlength\tabcolsep{4pt}
		\renewcommand{\arraystretch}{1.25}
		\begin{tabular}{|c|c|c|c|} \hline
			$\gamma_t$ & $\tau$ & CSR(MPS) & CSR(MPCA) \\ \hline
			\multirow{4}{*}{0.5} & 0.9 & 100 & 100 \\ \cline{2-4}
			& 0.8 & 100 & 100 \\ \cline{2-4}
			& 0.75 & 100 & 99.86 $\pm$ 0.44 \\ \cline{2-4}
			& 0.65 & 100 & 100 \\ \hline
			
			\multirow{4}{*}{0.2} & 0.9 & 100 & 98.02 $\pm$ 2.87 \\ \cline{2-4}
			& 0.8 & 100 & 97.59 $\pm$ 3.25 \\ \cline{2-4}
			& 0.75 & 99.04 $\pm$ 2.47 & 97.07 $\pm$ 4.81 \\ \cline{2-4}
			& 0.65 & 100 & 96.12 $\pm$ 2.94 \\ \hline
			
			\multirow{4}{*}{0.1} & 0.9 & 99.61 $\pm$ 0.98 & 94.15 $\pm$ 6.02 \\ \cline{2-4}
			& 0.8 & 98.29 $\pm$ 2.84 & 95.54 $\pm$ 5.35 \\ \cline{2-4}
			& 0.75 & 97.83 $\pm$ 3.67 & 91.92 $\pm$ 6.06 \\ \cline{2-4}
			& 0.65 & 99.15 $\pm$ 1.65 & 91.31 $\pm$ 4.47 \\ \hline
			
			\multirow{4}{*}{0.05} & 0.9 & 91.25 $\pm$ 7.29 & 80.29 $\pm$ 13.05 \\ \cline{2-4}
			& 0.8 & 96.99 $\pm$ 4.38 & 85.11 $\pm$ 11.45 \\ \cline{2-4}
			& 0.75 & 92.79 $\pm$ 6.05 & 82.7 $\pm$ 13.5 \\ \cline{2-4}
			& 0.65 & 94.78 $\pm$ 5.41 & 86.86 $\pm$ 6.12 \\ \hline
		\end{tabular}
	}
	\parbox{.45\linewidth}{
		\centering
		\caption{Binary Classification results (percent) for {\it EYFDB}. Faces with labels $3$ and $25$ were used.}
		\label{table:class-EYFDB}
		\setlength\tabcolsep{4pt}
		\renewcommand{\arraystretch}{1.25}
		\begin{tabular}{|c|c|c|c|} \hline
			$\gamma_t$ & $\tau$ & CSR(MPS) & CSR(MPCA) \\ \hline
			\multirow{4}{*}{0.5} & 0.9 & 99.25 $\pm$ 1.23 & 98.32 $\pm$ 2.16 \\ \cline{2-4}
			& 0.8 & 99.81 $\pm$ 0.51 & 99.34 $\pm$ 0.99 \\ \cline{2-4}
			& 0.75 & 99.84 $\pm$ 0.47 & 99.38 $\pm$ 0.99 \\ \cline{2-4}
			& 0.65 & 99.84 $\pm$ 0.47 & 99.49 $\pm$ 0.74 \\ \hline
			
			\multirow{4}{*}{0.2} & 0.9 & 97.51 $\pm$ 2.26 & 94.42 $\pm$ 3.92 \\ \cline{2-4}
			& 0.8 & 98.12 $\pm$ 1.57 & 97.67 $\pm$ 1.78 \\ \cline{2-4}
			& 0.75 & 98.45 $\pm$ 1.36 & 97.65 $\pm$ 3.02 \\ \cline{2-4}
			& 0.65 & 98.7 $\pm$ 1.05 & 97.16 $\pm$ 3.05 \\ \hline
			
			\multirow{4}{*}{0.1} & 0.9 & 95.07 $\pm$ 3.83 & 90.06 $\pm$ 4.72 \\ \cline{2-4}
			& 0.8 & 95.17 $\pm$ 3.88 & 91.29 $\pm$ 6.02 \\ \cline{2-4}
			& 0.75 & 95.4 $\pm$ 3.81 & 94.27 $\pm$ 5.78 \\ \cline{2-4}
			& 0.65 & 95.26 $\pm$ 3.88 & 90.5 $\pm$ 10.3 \\ \hline
			
			\multirow{4}{*}{0.05} & 0.9 & 88.31 $\pm$ 8.48 & 81.41 $\pm$ 9.64 \\ \cline{2-4}
			& 0.8 & 88.34 $\pm$ 8.86 & 84.14 $\pm$ 9.72 \\ \cline{2-4}
			& 0.75 & 88.52 $\pm$ 8.48 & 84.94 $\pm$ 8.03 \\ \cline{2-4}
			& 0.65 & 88.28 $\pm$ 9.44 & 82.19 $\pm$ 13.72 \\ \hline
		\end{tabular}
	}
\end{table}

\begin{table}[h t]
	\centering
	\caption{Binary Classification results (percent) for {\it MNIST}. Digits $5$ and $8$ were used.}
	\label{table:class-MNIST}
	\setlength\tabcolsep{4pt}
	\renewcommand{\arraystretch}{1.25}
	\begin{tabular}{|c|c|c|c|} \hline
		$\gamma_t$ & $\tau$ & CSR(MPS) & CSR(MPCA) \\ \hline
		\multirow{4}{*}{0.5} & 0.9 & 98.69 $\pm$ 0.09 & 98.92 $\pm$ 0.14 \\ \cline{2-4}
		& 0.8 & 98.83 $\pm$ 0.08 & 99 $\pm$ 0.08 \\ \cline{2-4}
		& 0.75 & 98.81 $\pm$ 0.1 & 98.95 $\pm$ 0.13 \\ \cline{2-4}
		& 0.65 & 98.77 $\pm$ 0.12 & 98.82 $\pm$ 0.08 \\ \hline
		
		\multirow{4}{*}{0.2} & 0.9 & 98.02 $\pm$ 0.11 & 98.3 $\pm$ 0.05 \\ \cline{2-4}
		& 0.8 & 98.1 $\pm$ 0.12 & 98.49 $\pm$ 0.1 \\ \cline{2-4}
		& 0.75 & 98.3 $\pm$ 0.07 & 98.45 $\pm$ 0.1 \\ \cline{2-4}
		& 0.65 & 98.27 $\pm$ 0.14 & 98.34 $\pm$ 0.13 \\ \hline
		
		\multirow{4}{*}{0.1} & 0.9 & 97.21 $\pm$ 0.22 & 97.63 $\pm$ 0.15 \\ \cline{2-4}
		& 0.8 & 97.4 $\pm$ 0.15 & 97.9 $\pm$ 0.17 \\ \cline{2-4}
		& 0.75 & 97.65 $\pm$ 0.18 & 97.85 $\pm$ 0.12 \\ \cline{2-4}
		& 0.65 & 97.67 $\pm$ 0.13 & 97.71 $\pm$ 0.15 \\ \hline
		
		\multirow{4}{*}{0.05} & 0.9 & 96.39 $\pm$ 0.33 & 96.7 $\pm$ 0.34 \\ \cline{2-4}
		& 0.8 & 96.61 $\pm$ 0.22 & 97.04 $\pm$ 0.14 \\ \cline{2-4}
		& 0.75 & 96.69 $\pm$ 0.27 & 97.2 $\pm$ 0.21 \\ \cline{2-4}
		& 0.65 & 96.93 $\pm$ 0.23 & 97.15 $\pm$ 0.22 \\ \hline
	\end{tabular}
\end{table}

%%%%%%%%%%%%%%%%%%%%%%%%%%%%%%%%%%%%%%%%%%%
% ROBUST TENSOR PCA
%%%%%%%%%%%%%%%%%%%%%%%%%%%%%%%%%%%%%%%%%%%

\section{Robust Tensor PCA}
\label{sec:RobustVersions}
An intrinsic limitation of both vector and tensor based dimensionality reduction methods is the sensitivity to the presence of outliers as the current decompositions focus on getting the best approximation to the tensor by minimizing the Frobenius norm of the error. In practice, the underlying tensor data is often low-rank, even though the actual data may not be due to outliers and arbitrary errors. It is thus possible to robustify tensor decompositions by reconstructing the low-rank part from its corrupted version. Recent years have seen the emergence of robust tensor decomposition methods \cite{goldfarb_robust_2014,Anandkumar_tensor_2016,Gu_robust_2014,inoue2009robust,inoue2009robustslrat,Huang_robust_2008,Pang_robust_2010}. The early attempts focused on solving the robust tensor problem using matrix methods, i.e. applying robust PCA (RPCA) to each matrix slice of the tensor or to the matrix obtained by flattening the tensor. However, such matrix methods ignore the tensor algebraic constraints such as the tensor rank which differs from the matrix rank constraints. Recently, the problem of low-rank robust tensor recovery has been addressed within CPD, HoSVD and t-SVD frameworks.
\subsection{Robust CPD}
The low-rank robust tensor recovery problem has been addressed in the context of CP model. Given an input tensor $\mathcal{X}=\mathcal{L}+\mathcal{S}$, the goal is to recover both $\mathcal{L}$ and $\mathcal{S}$, where $\mathcal{L}=\sum_{i=1}^{R}\sigma_{i}{\bf u}_{i}\otimes {\bf u}_{i}\otimes {\bf u}_{i}$ is a low-rank 3-mode tensor in the CP model and $\mathcal{S}$ is a sparse tensor. Anandkumar et al. \cite{Anandkumar_tensor_2016} proposed a robust tensor decomposition (RTD) for the sub-class of orthogonal low-rank tensors where $\langle {\bf u}_{i},{\bf u}_{j}\rangle=0$ for $i\neq j$. The proposed algorithm uses a non-convex iterative algorithm that maintains low rank and sparse estimates which are alternately updated. The low rank estimate $\mathcal{\hat{L}}$ is updated through the eigenvector computation of $\mathcal{X}-\mathcal{\hat{S}}$, and the sparse estimate is updated through thresholding of the residual $\mathcal{X}-\mathcal{\hat{L}}$. The algorithm proceeds in stages, $l=1,\ldots,R$, where $R$ is the target rank of the low rank estimate. In the $l$th stage, alternating steps of low rank projection to the l-rank space $P_{l}(\cdot)$ and hard thresholding of the residual are considered and the low rank projection is obtained through a gradient ascent method. The convergence of this algorithm is proven for rank-R orthogonal tensor $\mathcal{L}$ with block sparse corruption tensor $\mathcal{S}$.
\subsection{Robust HoSVD}
In the context of HoSVD, robust low-rank tensor recovery methods that rely on principal component pursuit (PCP) have been proposed \cite{goldfarb_robust_2014}. This method, referred to as Higher-order robust PCA, is a direct application of RPCA and defines the rank of the tensor based on the Tucker-rank (Trank). The corresponding tensor PCP optimization problem is
\begin{equation}
min_{\mathcal{L},\mathcal{S}} Trank(\mathcal{L})+\lambda\|\mathcal{S}\|_{0}, \hspace{0.2cm} s.t. \hspace{0.2cm}\mathcal{L}+\mathcal{S}=\mathcal{X}.
\end{equation}
This problem is NP-hard similar to PCP and can be convexified by replacing Trank($\mathcal{L}$) with the convex surrogate $CTrank(\mathcal{L})$ and $\|\mathcal{S}\|_{0}$ with $\|\mathcal{S}\|_{1}$:
\begin{equation}
min_{\mathcal{L},\mathcal{S}}CTrank(\mathcal{L})+\lambda\|\mathcal{S}\|_{1}, \hspace{0.2cm} s.t. \hspace{0.2cm} \mathcal{L}+\mathcal{S}=\mathcal{X}.
\label{eq:RPCAeq}
\end{equation}
Goldfarb and Qin \cite{goldfarb_robust_2014} considered variations of this problem based on different definitions of the tensor rank. In the first model, the Singleton model, the tensor rank regularization term is the sum of the $d$ nuclear norms of the mode-i unfoldings, i.e. $CTrank(\mathcal{L})=\sum_{i=1}^{d}\|L^{(i)}\|_{*}$. This definition of Ctrank leads to the following optimization problem:
\begin{equation}
min_{\mathcal{L},\mathcal{S}} \sum_{i=1}^{d}\|L^{(i)}\|_{*}+\lambda\|\mathcal{S}\|_{1}, \hspace{0.2cm} s.t. \hspace{0.2cm} \mathcal{L}+\mathcal{S}=\mathcal{X}.
\end{equation}
This problem can be solved using an alternating direction augmented Lagrangian (ADAL) method.

The second model, Mixture model, requires the tensor to be the sum of a set of component tensors, each of which is low-rank in the corresponding mode, i.e. $\mathcal{L}=\sum_{i=1}^{d}\mathcal{L}_{i}$, where $L_{i}^{(i)}$ is a low-rank matrix for each $i$. This is a relaxed version of the Singleton model which requires tensor to be low-rank in all modes simultaneously. This definition of the tensor rank leads to the following convex optimization problem
\begin{equation}
min_{\mathcal{L},\mathcal{S}} \sum_{i=1}^{d}\|L_{i}^{i}\|_{*}+\lambda\|\mathcal{S}\|_{1}, \hspace{0.2cm} s.t. \hspace{0.2cm} \sum_{i=1}^{N}\mathcal{L}_{i}+\mathcal{S}=\mathcal{X}.
\end{equation}
This is a more difficult optimization problem to solve than the Singleton model, but can be solved using an inexact ADAL algorithm.

Gu et al. \cite{Gu_robust_2014} extended Goldfarb's framework to the case that the low-rank tensor is corrupted both by a sparse corruption tensor as well as a dense noise tensor, i.e. $\mathcal{X}=\mathcal{L}+\mathcal{S}+\mathcal{E}$ where $\mathcal{L}$ is a low-rank tensor, $\mathcal{S}$ is a sparse corruption tensor, with the locations of nonzero entries unknown, and the magnitudes of the nonzero entries can be arbitrarily large, and $\mathcal{E}$ is a noise tensor with i.i.d Gaussian entries. A convex optimization problem is proposed to estimate the low-rank and sparse tensors simultaneously:
\begin{equation}
argmin_{\mathcal{L},\mathcal{S}} \|\mathcal{X}-\mathcal{L}-\mathcal{S}\|^{2}+\lambda \|\mathcal{L}\|_{S_{1}}+\mu \|\mathcal{S}\|_{1},
\end{equation}
where $\|\cdot\|_{S_1}$ is tensor Schatten-1 norm, $\|\cdot\|_{1}$ is entry-wise $\ell_{1}$ norm of tensors. Instead of considering this observation model, the authors consider a more general linear observation model and obtain the estimation error bounds on each tensor, i.e. the low-rank and the sparse tensor. This equivalent problem is solved using ADMM.

\subsection{Robust t-SVD}
More recently, tensor RPCA problem has been extended to the t-SVD and its induced tensor tubal rank and tensor nuclear norm \cite{lu2016tensor}. The low-rank tensor recovery problem is posed by minimizing the low tubal rank component instead of the sum of the nuclear norms of the unfolded matrices. This results in a convex optimization problem. Lu et al. \cite{lu2016tensor} prove that under certain incoherence conditions, the solution to \ref{eq:RPCAeq} with convex surrogate Tucker rank replaced by the tubal rank perfectly recovers the low-rank and sparse components, provided that the tubal rank of $\mathcal{L}$ is not too large and $\mathcal{S}$ is reasonably sparse. Zhou and Feng \cite{zhou2017outlier} address the same problem using an outlier-robust tensor PCA (OR-TPCA) method for simultaneous low-rank tensor recovery and outlier detection. OR-TPCA recovers the low-rank tensor component by minimizing the tubal rank of $\mathcal{L}$ and the $\ell_{2,1}$ norm of $\mathcal{S}$. Similar to \cite{lu2016tensor}, exact subspace recovery guarantees for tensor column-incoherence conditions are given.

The main advantage of t-SVD based robust tensor recovery methods with respect to robust HoSVD methods is the use of the tensor nuclear norm instead of the Tucker rank in the optimization problem. It can be shown that the tensor nuclear norm is equivalent to the nuclear norm of the block circulant matrix which preserves relationships among modes and better depicts the low-rank structure of the tensor. Moreover, minimizing the tensor nuclear norm of $\mathcal{L}$ is equivalent to recovering the underlying low-rank structure of each frontal slice in the Fourier domain \cite{lu2016tensor}.

\section{Conclusion}
\label{sec:Conc}
%\input{tensorPCA_conclusions}
%tensorPCA conclusions;
In this paper we provided an overview of the main tensor decomposition methods for data reduction and compared their performance for different size tensor data. We also introduced tensor PCA methods and their extensions to learning and robust low-rank tensor recovery applications. Of particular note, the empirical results in Section~\ref{sec:TensorDecomps} illustrate that the most often used tensor decomposition methods depend heavily on the characteristics of the data (i.e., variation across the slices, the number of modes, data size, \dots) and thus point to the need for improved decomposition algorithms and methods that combine the advantages of different existing methods for larger datasets. Recent research has focused on techniques such as the block term decompositions (BTDs) \cite{sorber2013optimization,de2008decompositions,de2008decompositions3,de2011blind} in order to achieve such results.  BTDs admit the modeling of more complex data structures and represent a given tensor in terms of low rank factors
that are not necessarily of rank one.  This enhances the potential for modeling more general phenomena and can be seen as a combination of the Tucker and CP decompositions.

The availability of flexible and computationally efficient tensor representation tools will also have an impact on the field of supervised and unsupervised tensor subspace learning. Even though the focus of this paper has been mostly on tensor data reduction methods and not learning approaches, it is important to note that recent years have seen a growth in learning algorithms for tensor type data. Some of these are simple extensions of subspace learning methods such as LDA for vector type data to tensors \cite{tao2005supervised,li2008discriminant} whereas others extend manifold learning \cite{mordohai2010dimensionality,guo2016tensor,zhong2014low,wu2012learning} to tensor type data for computer vision applications.

As the dimensionality of tensor type data increases, there will also be a growing need for hierarchical tensor decomposition methods for both visualization and representational purposes.  In particular, in
the area of large volumetric data visualization, tensor based multiresolution hierarchical methods such as TAMRESH \cite{suter2013tamresh} have been considered, and in the area of data reduction and denoising multiscale HoSVD methods have been proposed \cite{ozdemir2015locally,ozdemir2016multiscale,ozdemir2016multiscale2,ozdemir2017multiscale}. This increase in tensor data dimensionality will also require the development of parallel, distributed implementations of the different decomposition methods \cite{sidiropoulos2012multi,sidiropoulos2014parallel}. Different strategies for the SVD of large-scale matrices encountered both for vector and tensor type data \cite{Halko_Finding_2011,Iwen_Distributed_2016,cichocki2014era,caiafa2010generalizing,huy2011parafac} have already been considered for efficient implementation of the decomposition methods. 

\bibliographystyle{IEEEbib}
\bibliography{TensorPCA,references_alp}

%\pagebreak

\vspace{4\baselineskip}
\parpic{\includegraphics[width=1in,clip,keepaspectratio]{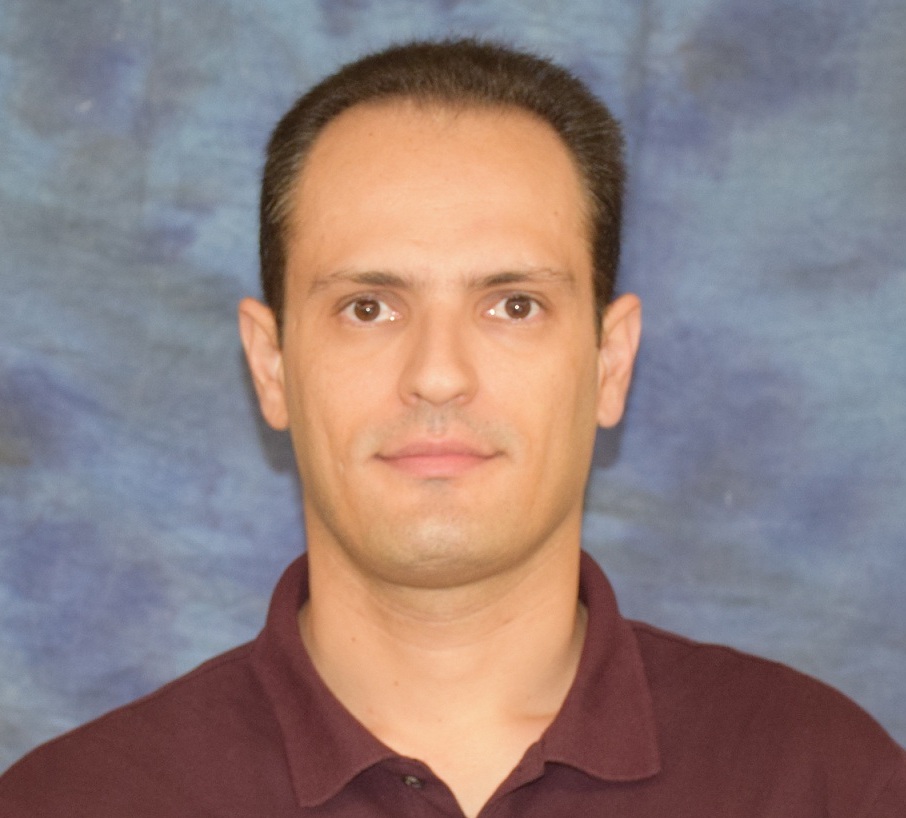}}
\noindent \textbf{Ali Zare} received his B.Sc. in Electrical Engineering and his M.Sc. in Electrical Engineering: Communication Systems from Shiraz University, Shiraz, Iran, in 2006 and 2011, respectively. He is a Ph.D. student in Computational Mathematics, Science and Engineering at Michigan State University. His research interests include signal processing, machine learning and processing of higher-order data.\\ \\

\parpic{\includegraphics[width=1in,clip,keepaspectratio]{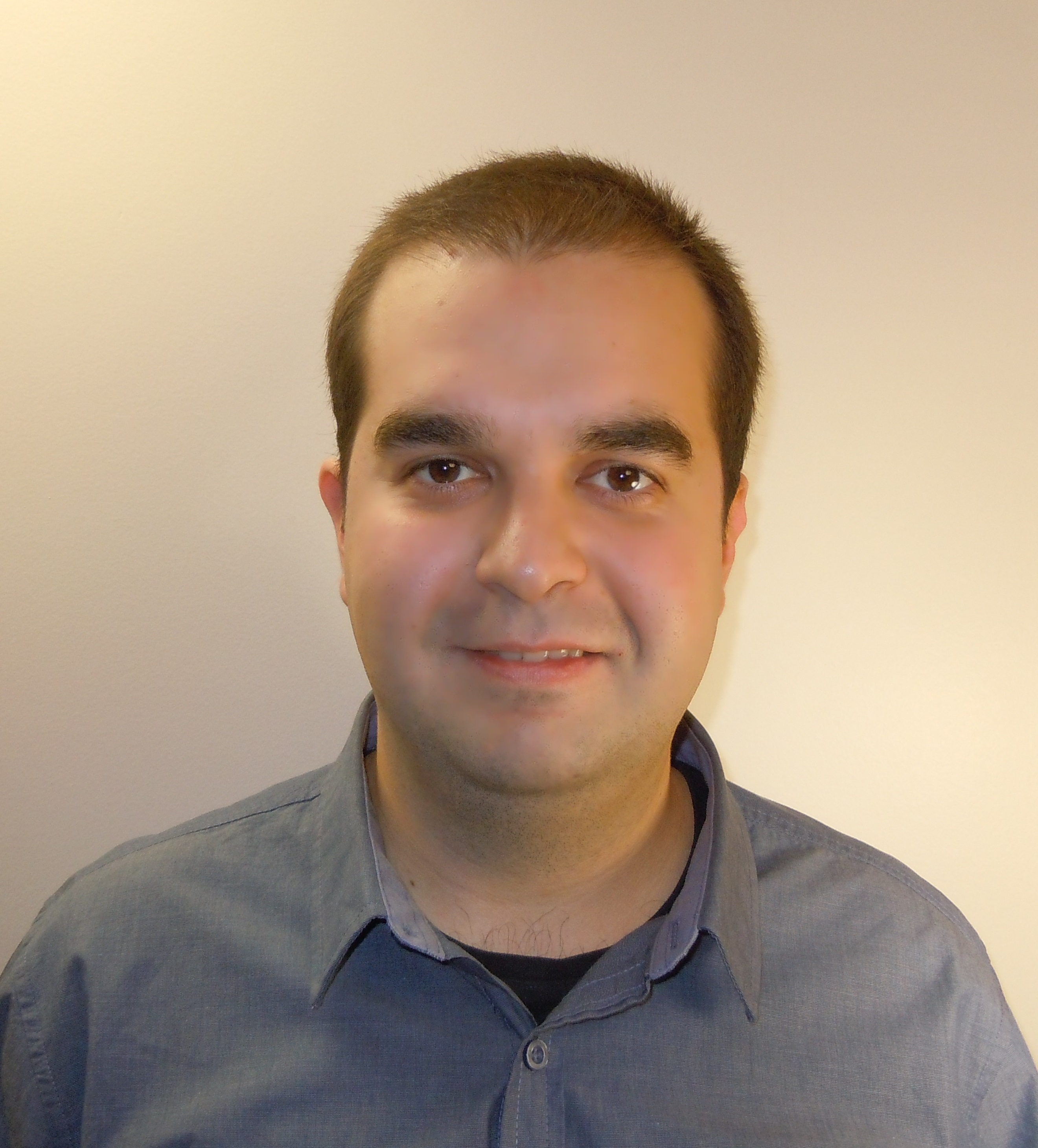}}
\noindent \textbf{Alp Ozdemir} received the B.S. degree in Electrical and Electronics Engineering from Bogazici University, Istanbul in 2011 and the M.S. degree in Biomedical Engineering from Bogazici University, Istanbul in 2013. He received Ph.D. degree in Electrical and Computer Engineering at Michigan State University, East Lansing in 2017. His research focuses on signal processing, image processing and machine learning with applications to tensors.\\ \\

\parpic{\includegraphics[width=1in,clip,keepaspectratio]{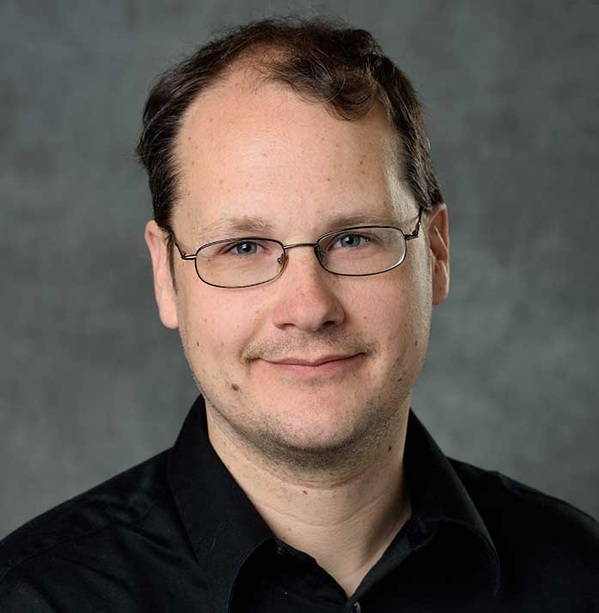}}
\noindent \textbf{Mark Iwen} earned a Ph.D. in 2008 from the University of Michigan in Applied and Interdisciplinary Mathematics. From September 2008 through August 2010 he was a postdoctoral fellow at the Institute for Mathematics and its Applications (IMA), and then moved to Duke University as a visiting assistant professor from September of 2010 until August 2013. He has been an assistant professor at Michigan State since the fall of 2013.	His research interests include signal processing, computational harmonic analysis, and algorithms for the analysis of large and high dimensional data sets.\\ \\

\parpic{\includegraphics[width=1in,clip,keepaspectratio]{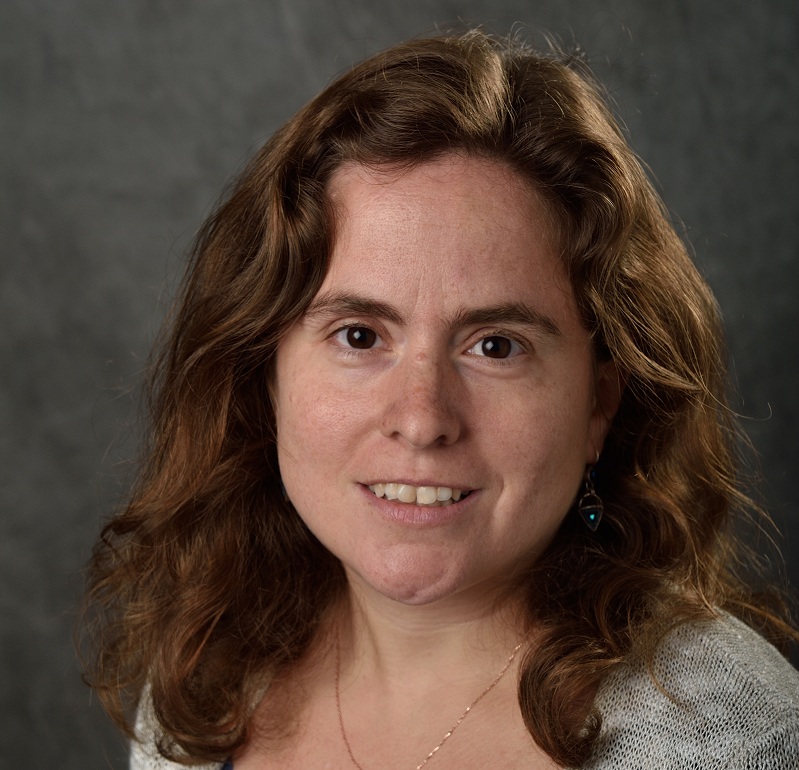}}
\noindent \textbf{Selin Aviyente} Selin Aviyente received her B.S. degree with high honors in Electrical and Electronics engineering from Bogazici University, Istanbul in 1997. She received her M.S. and Ph.D. degrees, both in Electrical Engineering: Systems, from the University of Michigan, Ann Arbor, in 1999 and 2002,
respectively. She joined the Department of Electrical and Computer
Engineering at Michigan State University in 2002, where she is currently a
Professor. Her research focuses on statistical signal processing,
higher-order data representations and complex network analysis with
applications to biological signals.  She has authored more than 150
peer-reviewed journal and conference papers. She is the recipient of a
2005 Withrow Teaching Excellence Award and a 2008 NSF CAREER Award. She is
currently serving on several technical committees of IEEE Signal
Processing Society including the Signal Processing Theory and Methods and
Bio-imaging and Signal Processing Technical Committees. She is a Senior
Area Editor for IEEE Transactions on Signal Processing and an Associate
Editor for IEEE Transactions on Signal and Information Processing in
Networks.

\end{document}